\documentclass[a4paper,11pt]{article}
\usepackage{jheppub} 
\usepackage{lineno}
\newcommand{\rmd}{\mathrm{d}}
\usepackage{eufrak}
\usepackage{amsmath}
\usepackage{mathtools}

\title{\boldmath Critical JT Gravity}

\author{Alicia Castro}
\affiliation{IMAPP, Radboud University,\\
6525 AJ Nijmegen, Netherlands}

\emailAdd{alicia.castro@science.ru.nl}
\abstract{In this paper, we investigate a critical behavior of JT gravity, a model of two-dimensional quantum gravity on constant negatively curved spacetimes. Our approach involves using techniques from random maps to investigate the generating function of Weil--Petersson volumes, which count random hyperbolic surfaces with defects. The defects are weighted geodesic boundaries, and criticality is reached by tuning the weights to the regime where macroscopic holes emerge in the hyperbolic surface, namely \textit{non-generic criticality}. We analyze the impact of this critical regime on some universal features, such as its density of states. We present a family of models that interpolates between systems with $\rho_0(E)\sim\sqrt{E-E_0}$ and $\rho_0(E)\sim (E-E_0)^{3/2}$, which are commonly found in models of JT gravity coupled to dynamical end-of-the-world and FZZT branes, and give a precise definition of what this phase transition means from the random geometry point of view.}

\begin{document}
\maketitle
\flushbottom

\section{Introduction}

Quantizing gravity remains one of the most challenging and unresolved problems in physics. The traditional methods for quantizing physical systems fail to apply when it comes to gravity. For instance, when using the path integral formalism, one needs to integrate over the space of all possible 4-dimensional spacetimes governed by Einstein-Hilbert's action. However, this space is not yet completely understood mathematically, nor are the physical constraints, such as topology change and Lorentzian versus Euclidean signatures \cite{Ambjorn:1998fd}, well-defined.  As a result, studying lower-dimensional toy models is often a useful tool for analyzing these issues.

Two-dimensional quantum gravity is a useful toy model from both the physical and mathematical perspectives. Physically, 2D quantum gravity allows us to study key features of gravity without having to deal with propagating degrees of freedom i.e.\ gravitons  Mathematically, the existence of conformal symmetry and discretization techniques in two dimensions make some aspects of 2D quantum gravity computable. In particular, the 2D Einstein-Hilbert action is topological, so the euclidean path integral reduces to counting geometries with a weight given by $e^{S_0\chi}$, where $\chi$ is the Euler characteristic and $S_0\sim\frac{1}{G}$ (see \cite{Budd:2022zry} for an updated review). Yet, the exact quantization of 2D gravity remains a compelling task.

Recent developments in JT gravity \cite{Jackiw:1984je,Teitelboim:1983ux}, a model of euclidean 2-dimensional quantum gravity on constant negatively curved spacetimes (the hyperbolic plane, which is the euclidean version of AdS$_2$), have proved potential for shedding light on this topic due to its simplicity and rich physical content \cite{Mertens:2022irh}. One of the fascinating features of JT gravity is its connection to random geometries \cite{Witten:2020bvl}.  In particular, it has been demonstrated that the partition function of JT gravity can be split into two parts: one is proportional to volumes of moduli spaces of hyperbolic surfaces, while the other only involves information about the asymptotic behavior of AdS$_2$. The latter component can be interpreted in terms of a holographic correspondence, as it is the partition function of a quantum mechanical system given by the Schwarzian derivative effective action near the asymptotic boundary \cite{Maldacena:2016upp, Kitaev}. Furthermore, the Weil-Petersson volumes carry the geometrical information of the negatively-curved surface. In addition, studying the higher genus contributions to this partition function revealed its relation with yet another random geometry field, matrix models \cite{Saad:2019lba,Stanford:2019vob}. This allowed for significant progress in the field since the formal machinery of matrix models \cite{Eynard:2016yaa} can be applied to JT gravity \cite{Johnson:2019eik}.

\begin{figure}[htbp]
\centering
\includegraphics[width=.5\textwidth]{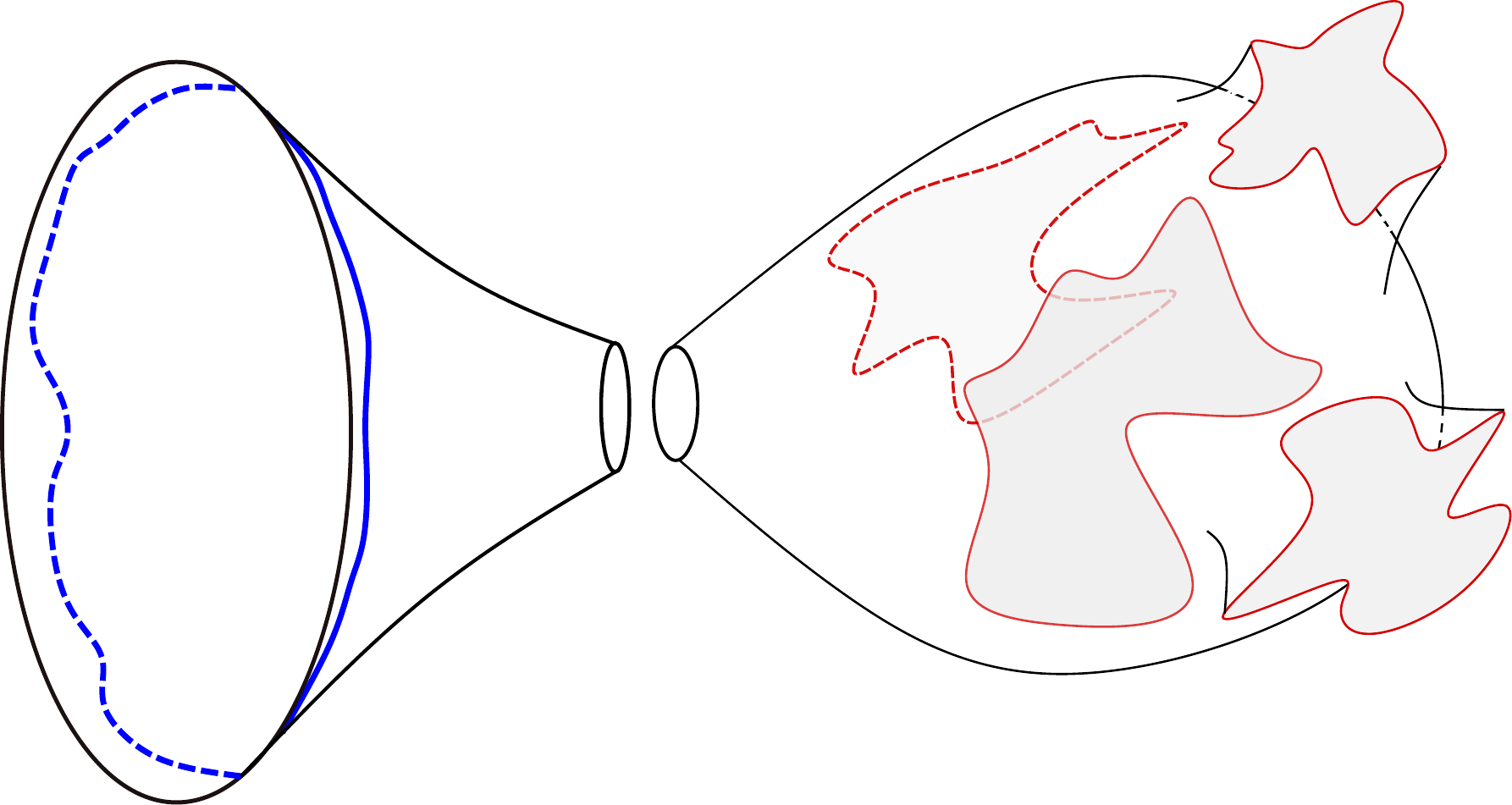}
\caption{Example of a configuration of critical JT gravity. The blue line represents the nearly AdS boundary and in red are the geodesics boundaries. In the non-generic critical regime, these boundaries become macroscopic.\label{fig:JT_picture_holes}}
\end{figure}

In this paper, we apply novel techniques from random geometry \cite{budd2023} to JT gravity. We begin by examining the generating function of Weil-Petersson volumes with non-trivial weights on its geodesic boundaries. This allows us to investigate the behavior of the JT gravity partition function, in particular, the regime where there is a non-analytical behavior of the disk function and density of states, which are indicative of non-generic criticality. Such critical regimes are reminiscent of the planar map case where there is a transition from a phase of surfaces with microscopic holes to a phase with macroscopic ones (see Figure \ref{fig:JT_picture_holes}). This translates, in the hyperbolic geometry case, to the proliferation of large geodesic boundaries. We argue the existence of these phases by looking at the hyperbolic $O(n)$ loop model. This phase transition is driven by a fine-tuning of the geodesic weights and has a significant impact on the density of states of the theory in the boundary. Our approach allows us to identify a family of models that interpolates between systems with $\rho_0(E)\sim\sqrt{E}$ and $\rho_0(E)\sim E^{3/2}$, which are commonly found in JT gravity coupled to dynamical end-of-the-world (EOW)\cite{Gao:2021uro} and FZZT branes \cite{Okuyama:2021Page}.\\
\\
This paper is organized as follows. In section \ref{section:Random_geometry} we introduce random maps and random hyperbolic surfaces as examples of random geometries, together with the notion of criticality. In section \ref{section_JT_random}, we introduce a dictionary between the vocabularies of random geometry and JT gravity using a common tool, the string equation. To clarify, we reproduce results on FZZT branes and conical defects in these terms. In Section \ref{Section_Criticality}, we define and analyze the different critical regimes of the string equation and we focus on its implication for the partition and disks functions and density of states in the non-generic critical phase. In section \ref{Section_Dilaton_gravity}, we construct the corresponding dilaton potential and analyze the thermodynamical properties of these spacetimes and their relation to previous studies using branes and matrix models. In Section \ref{section:SFF}, we compute the spectral form factor of this family of models and comment on its potential to reach a late-time plateau. Finally, in section \ref{Section_Discussion}, we give an outlook and discuss potential implications for JT gravity. 

\section{Random Geometry}\label{section:Random_geometry}
In order to make this paper as self-contained as possible, let us recall some terms that are useful to describe random geometries, for more details see \cite{Eynard:2016yaa} and \cite{Budd:2022zry} for an updated review. A \textbf{\textit{graph}} is a finite set of vertices, edges and an incidence map  $\left\{ v,e,\mathbb{I}\right\} $, $\mathbb{I}:e\rightarrow (v_1,v_2)$ that associates two vertices with an edge. A \textbf{\textit{genus-g map}} is an equivalence class of \textit{graphs} embedded in a surface of genus-g, such that its edges do not cross, modulo orientation-preserving homeomorphisms of the surface.  In particular, throughout this paper, when we talk about genus-$g$ maps, we will only consider maps with one marked edge (called the root), a defined orientation (an arrow that goes from one vertex to the other on the root edge), pointed (one distinguished vertex of the root) and an even number of edges per face. In the case $g=0$, these are called rooted pointed planar bipartite maps. A face of a genus-g map is a topological disk bounded by edges. The \textit{length of a face} is the sum of the lengths of its edges, where each edge has length $1$.


\subsection{Random maps}
Let $\mathcal{M}_{g,n}(L_1,\dots,L_n)$ be the set of genus-$g$ maps with $n$ faces of half-lengths\footnote{Recall, all faces have an even number of edges, so their length is $2L_i$.} $\{L_i\in \mathbb{N}\}_{1\leq i\leq n}$ and weights $\mathbf{q}=(q(L_i)\geq 0)$ per face of half-length $L_i$ and with $q(0)$ is the weight per vertex which we will usually set to $1$. These weights can be understood as a decoration of the map which can be associated with a coupling to matter. The cardinality of this set is the total number of Boltzmann genus-$g$ maps up to orientation-preserving automorphisms that preserve the face labels
\begin{equation}
    ||\mathcal{M}_{g,n}(L_1,\dots,L_n)||=\sum_{\mathfrak{m}\in\mathcal{M}_{g,n}({L_i})}\frac{1}{\mathrm{Aut}(\mathfrak{m})}.
\end{equation}
For fixed values of $g$ and $n$, these are nothing more than positive real numbers and they can be encoded in a generating function
\begin{equation}
  F_g[\mathbf{q}]= \sum_{k=1}^\infty\frac{1}{k!} \sum_{L_1=0}^\infty q(L_1) \cdots \sum_{L_k=0}^\infty q(L_k) q(0)^{2-2g-k+\sum_i L_i} ||\mathcal{M}_{g,k}(L_1,\dots,L_k)||.
\end{equation}
where the powers of $q(0)$ account for the number of vertices with weight $q(0)$ in a map of genus-$g$ with $k$ faces of half-length $\sum_i L_i$ according to Euler's characteristic. Therefore, we can write  
\begin{equation}
     ||\mathcal{M}_{g,n}(L_1,\dots,L_n)||=q(0)^{-2+2g+n-\sum_i L_i}\left.\frac{\partial^n F_g[\mathbf{q}]}{\partial q(L_1)\dots \partial q(L_n)}\right|_{\mathbf{q}=0}.
\end{equation}
$F_g[\mathbf{q}]$ is called the \textit{partition function of genus-g maps}. It can be observed that the weights $q(L_k)$ control the likelihood of having faces of length $L_k$ in the map, e.g. if $\mathbf{q}=(q(0)=1,q(1)=0,q(2)\neq 0,q(3)=0,0,0,\dots)$, then only faces of length $2\cdot2$ are allowed and the genus-$g$ map is a quadrangulation of a genus-$g$ surface. Coming back to the generating function $F_g[\mathbf{q}]$, in order to analyse it, it is convenient to write it in terms of another power series, the \textit{string equation} 
\begin{equation}
   \tilde{x}(u)=u-\sum_{k=0}^\infty q(k)\;\frac{1}{2}\binom{2k}{k} u^k,\label{eq:string_eq_maps}
\end{equation}
where $q(k)\coloneqq q(L=k)$ with $k\in \mathbb{N}$. In particular, the genus-$0$ generating function is proven to be \cite{Tutte1968OnTE,Eynard:2016yaa}\footnote{For $g>0$, check \cite{Eynard:2016yaa}.}
\begin{align}
    F_0 &= \frac{1}{2}\int_{0}^{E_\mathbf{q}} \!\!\!\rmd u\,\frac{1}{u} \left(\tilde{x}(u)^2 -(u-q(0))^2 \mathbb{I}_{u<q(0)} \right),\label{eq:genfun0beta}
\end{align}
where $E_\mathbf{q}>0$ is the smallest solution to the string equation $\tilde{x}(E_\mathbf{q})=0$. Thus, $E_\mathbf{q}$ is an implicit function of the weight $\mathbf{q}$. The weight is said to be \textit{admissible} if such a solution exists. Furthermore, another useful quantities we are interested are the \textit{moment} $M_0$, which is defined 
\begin{equation}
M_0=\left(\frac{\partial E_\mathbf{q}}{\partial q(0)}\right)^{-1}=\frac{\partial \tilde{x}}{\partial u}(E_\mathbf{q}),
\end{equation}
and the disk function, which corresponds to the generating function of genus-$0$ maps with one marked face and an arbitrary number of unmarked faces, operationally, this is 
\begin{equation}
    W_\mathbf{q}(L) = \frac{\partial 	F_0[\mathbf{q}]}{\partial q(L)}.
\end{equation}
It can also be noted that
$\frac{\partial^2	F_0[\mathbf{q}]}{\partial q(0)\partial q(1)}=E_\mathbf{q}
$
which gives a geometrical meaning to $E_\mathbf{q}$, it is the generating function of rooted, pointed planar maps.\\
By this point, all our expressions are formal power series, but in order to talk about random geometry and random maps, we need a well-defined probability measure. This can be also understood as a well-defined canonical way of sampling a map uniformly. In particular, we would like to define a probability density measure on the space of planar maps with $k$ faces with weight $\mathbf{q}$. This can be done using the generating function in the following way   
\begin{equation}
    p\left(L_1, ... , L_k\right) = \frac{1}{k!} \frac{\prod_{i=1}^{k} q(L_i)}{F_0[\mathbf{q}]}.
\label{eq:prob_density_boltzmann}
\end{equation}
Note that this is only well defined when $\mathbf{q}$ is such that $F_0[\mathbf{q}]<\infty$. This is satisfied iff the weight $\mathbf{q}$ is admissible, i.e.\, if the string equation has a solution $E_\mathbf{q}>0$. In this way, given an admissible weight, a \textbf{\textit{Boltzmann planar map}} is defined as a random planar map with $k$-faces of random lengths sampled with probability \eqref{eq:prob_density_boltzmann}.
\subsection{Critical maps and the $O(n)$ loop model}
\begin{figure}[htbp]
\centering
\includegraphics[width=.8\textwidth]{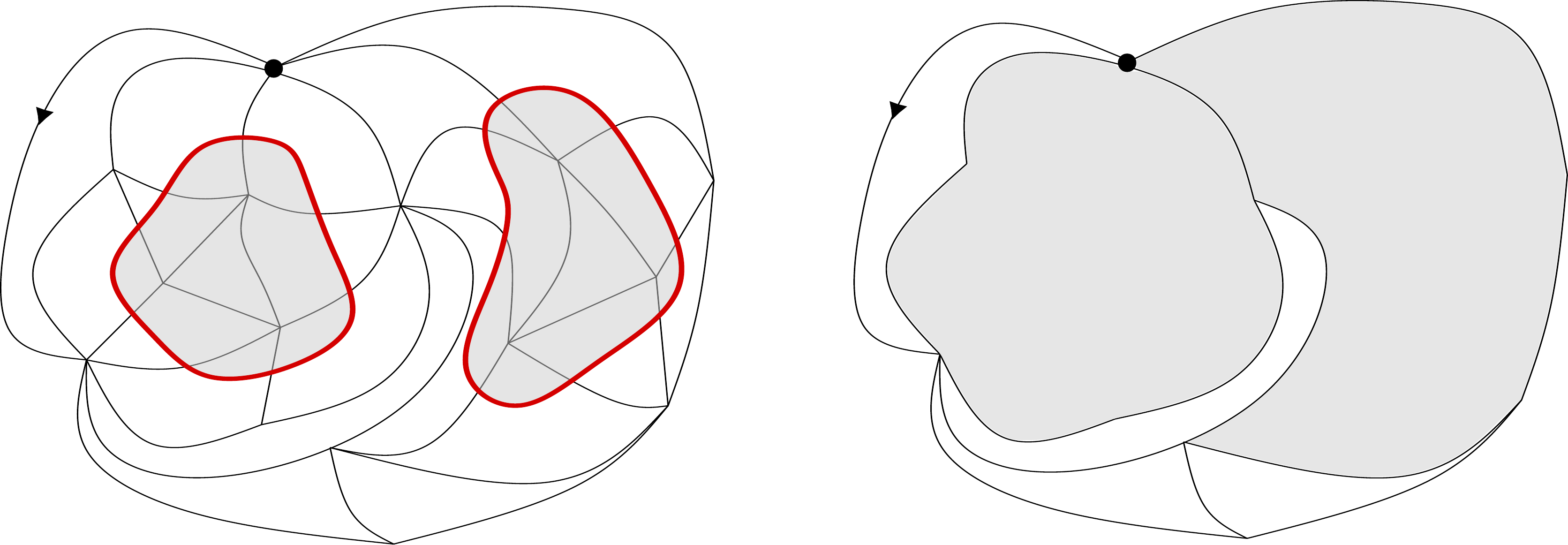}
\caption{\textbf{Left}: Illustrative example of a $O(n)$ model configuration with two non-intersecting loops in red. The interiors of the loops are colored grey. \textbf{Right}: The corresponding external gasket where the marked faces, or holes, are colored in gray.\label{fig:holes_maps}}
\end{figure}

Now that we introduced Boltzmann planar maps, we can study more of their characteristics. In particular, the main point we want to make is that the string equation is the central building block to construct all we have done so far. Consider the solution to the string equation $\tilde{x}(E_\mathbf{q})/E_\mathbf{q}=0$ and its first derivative $\tilde{x}'(E_\mathbf{q})$,
\begin{equation}
    1-E_\mathbf{q}^{-1}=\sum_{k=1}^\infty q(k)\;\frac{1}{2}\binom{2k}{k} E_\mathbf{q}^{k-1}, \hspace{1cm} \tilde{x}'(E_\mathbf{q})=1-\sum_{k=1}^\infty q(k)\;\frac{k}{2}\binom{2k}{k} E_\mathbf{q}^{k-1}.
\end{equation}
Given that $q(k),E_\mathbf{q}>0$, we can conclude that $0\leq \tilde{x}'(E_\mathbf{q})<1$.
In the mathematics literature, if an admissible weight is such that $0<\tilde{x}'(E_\mathbf{q})$, it is called \textit{subcritical}, and if $\tilde{x}'(E_\mathbf{q})=0$, the weight is called \textbf{\textit{critical}}. Furthermore, if a weight is critical, it can be \textit{generic critical} if the string equation behaves as $\tilde{x}(u) \sim (E_\mathbf{q}-u)^2$ for $u\lesssim E_\mathbf{q}$ and \textbf{\textit{non-generic critical}} if the string equation behaves as $\tilde{x}(u)\sim (E_\mathbf{q}-u)^{\alpha-1/2}$ with $3/2<\alpha<5/2$ for $u\lesssim E_\mathbf{q}$ \cite{10.2307/25450028}. In this last case, $E_\mathbf{q}>0$ is the radius of convergence of the formal power series $\tilde{x}(u)$.
\begin{table}[h]
    \centering
    \begin{tabular}{|c|c|}
    \hline
        Weight $\mathbf{q}$ & String equation \\
    \hline     
         Subcritical & $\tilde{x}'(E_\mathbf{q})> 0$ 
         \\
         Generic critical & $\tilde{x}'(E_\mathbf{q})=0, \hspace{2mm} \tilde{x}(u) \sim (E_\mathbf{q}-u)^2$
         \\
         Non-generic critical & $\tilde{x}'(E_\mathbf{q})=0, \hspace{2mm} \tilde{x}(u)\sim (E_\mathbf{q}-u)^{\alpha-1/2}$\\
         \hline
    \end{tabular}
    \caption{Types of admissible weights.}
    \label{tab:critical}
\end{table}\\
In the non-generic critical case, it is proven that the weight must behave as $\mathbf{q}(k)\sim h_1^{k}k^{-\alpha}$ for some $h_1>0$ and $k\rightarrow\infty$ \cite{legall,buddcurien}. The geometrical meaning of this slow decay of critical weights is the proliferation of large faces $k\gg 1$ (See right part of Figure \ref{fig:holes_maps}). 
To further understand this, we will follow the reasoning used to draw a relation between these maps and the $O(n)$ loop model in \cite{Borot:2011kg,2018arXiv180902012B}. In that case, one considers a planar map dressed with a loop configuration which is
a collection of disjoint loops drawn on the planar map each of arbitrary length $2k$, where the parameter $n$ controls the weight per loop. The external gasket $E(n,\Gamma)$ is then obtained by removing all edges and vertices situated within the interior of the loops in $\Gamma$, as well as those intersected by them, from the planar map, these faces of the exterior gasket are called \textit{holes} (see Figure \ref{fig:holes_maps}). The random gasket configurations $O(n)$ loop configurations with $n=2 \sin\left((\pi(\alpha-1/2)\right)$ are related to non-generic critical Boltzmann planar maps for $\alpha\in (3/2,5/2)$ \cite{Borot_2012}. The construction consists on substituting the faces touched and contained on each loop by holes with weights assigned consistently with the disk functions relation $W^{(n)}_\mathbf{q}(k)=W_\mathbf{\hat{q}}(k)$, where $W^{(n)}_\mathbf{q}(k)$ is the disk function of the $ O(n)$ loop model with weights $\mathbf{q}$, and $\hat{\mathbf{q}}(k)=\mathbf{q}(k)+n\ h_1^{k} W_\mathbf{\hat{q}}(k)$. We will make this clear by studying this explicit construction in the hyperbolic case.
\\
We will come back to this topic in the next subsection and elaborate more on the hyperbolic surfaces case, but interested readers can refer to \cite{Borot:2011kg} for proofs concerning the $O(n)$ loop model in the context of maps.



\subsection{Random hyperbolic surfaces}
Along the lines of random planar maps, we want to illustrate how to consistently define a random hyperbolic surface by constructing a probability measure in the space of hyperbolic surfaces of genus $g$ and $k$ geodesic boundaries, $\mathcal{M}^{WP}_{g,k}(b_1,\dots,b_k)$. The starting point is to write the `cardinality' of this space, this is given by the volume of the moduli space of $\mathcal{M}^{WP}_{g,k}(b_1,\dots,b_k)$, called Weil-Petersson volumes $V^{WP}_{g,k}(b_1,\dots,b_k)$ \cite{Mirzakhani}. The second step is to write a generating function of Weil-Petersson volumes of genus $g$ \cite{1996CMaPh.181..763K}
\begin{align}
	F^{WP}_g[\mu] &= \sum_{k=0}^\infty\frac{2^{2-2g-k}}{k!} \int_{0}^\infty \rmd \alpha_1 \mu(\alpha_1) \cdots \int_{0}^\infty \rmd \alpha_k \mu(\alpha_k) V^{WP}_{g,k}(\alpha_{1}\ldots,\alpha_{k}),
\end{align}
Note that similarly to the Boltzmann map case, $\mu(\ell_k)$ can be interpreted as the likelihood of having geodesic boundaries of length $\ell_k$, i.e.\ it is a weight\footnote{Another way to see this is to consider a weight $\mu(\ell)=x\tilde{\mu}(\ell)$ to make a clear distinction between the formal variable of the generating function, $x$, and the statistical weight $\tilde{\mu}$ in the sense of a non-negative valued label on the geodesic boundaries of the hyperbolic surface.}. In Section \ref{section_JT_random} we will explain the role that weights play in JT gravity with defects.\\
The generating function of Weil-Petersson volumes can be expressed in terms of the string equation 
\begin{equation}
   x(u)=u-\sum_{k=0}^\infty\frac{t_k+\gamma_k}{k!}u^k,\label{eq:string_eq_times}
\end{equation}
where 
\begin{equation}
\gamma_k=\frac{(-1)^k\pi^{2k-2}}{(k-1)!}\mathrm{1}_{k\geq 2}\label{gamma_k}
\end{equation}
are the KdV times \cite{1996CMaPh.181..763K, Mirzakhani} and are called the times of pure JT gravity in the physics literature \cite{Johnson:2019eik,Okuyama:2019xbv}, and
\begin{equation}
    t_k=\frac{4^{-k}}{k!}\int_0^\infty \rmd \ell \mu(\ell) \ell^{2k}\label{eq:def_general_JT_times}
\end{equation}
are the times associated with geodesic boundaries with labels $\mu(\ell)$. Note that for Boltzmann maps, we have $\gamma_k=0$.
 Equivalently, the string equation \eqref{eq:string_eq_times} can be written in the following form 
\begin{equation}
    x(u)=\frac{\sqrt{u}}{\pi}J_1\left(2\pi\sqrt{u}\right)-\int_0^\infty \rmd\ell \mu(\ell) I_0\left(\ell \sqrt{u}\right).\label{eq:general_string_eq_JT}
\end{equation}
The first term is well-known in the physics literature as \textit{pure JT gravity}, up to a change of sign $u\rightarrow -u$. The second term accounts for the presence of geodesic boundaries of weight $\mu(\ell)$ and it has been used in the physics literature couple JT gravity to conical defects and branes \cite{Turi:2020ale,Okuyama:2021FZZT,Gao:2021uro}.\\
In the rest of this section, we will restrict our study to the case $g=0$ for convenience, but the interested reader can check \cite{budd2022irreducible,budd2023_2} for $g>0$. The genus-$0$ generating function of Weil-Petersson volumes can be expressed in terms of the string equation in the same way as for Boltzmann planar maps, as well as the moment $M_0$ \cite{budd2022irreducible}
\begin{align}
    	F^{WP}_0[\mu] &= \frac{1}{4}\int_{0}^{E_\mu} \!\!\!\rmd u\, x(u)^2,\label{eq:genfun0}\\
    M^{WP}_0&=\frac{\partial x}{\partial u} (E_\mu)\label{eq:m0_wp}
\end{align}
where $E_\mu>0$ is the smallest solution to $x(E_\mu)=0$, and, in the physics literature, it is called the threshold energy $E_\mu=-E_0$.  This allows us to define a probability density measure on the space of hyperbolic genus-$0$ surfaces  
\begin{equation}
    p^{WP}\left(\ell_1, ... , \ell_k\right) = \frac{1}{k!} \frac{\prod_{i=1}^{k}  \mu(\ell_i)}{F^{WP}_0[\mu]}.
\label{eq:prob_density_WP}
\end{equation}
In this way, given an admissible weight $\mu$ such that $F^{WP}_0[\mu]<\infty$, we can call \textbf{\textit{random genus-$0$ hyperbolic surface with $k$-boundaries of random lengths}} to a surface sampled with probability \eqref{eq:prob_density_WP}.\\
In particular, once again, it is useful to introduce the disk function, this is the generating function of hyperbolic surfaces with one marked geodesic boundary of length $b$
\begin{align}
W_\mu(b) &=\frac{\delta F^{WP}
_0}{\delta\mu(b)}=\sum_{k=2}^\infty\frac{2^{2-(k+1)}}{k!}\int_0^\infty \rmd \ell_1
\mu(\ell_1)...\int_0^\infty \rmd \ell_k
\mu(\ell_k)V^{WP}_{0,k+1}(b,\ell_1,...,\ell_k)\\
   &= -\frac{1}{2}\int_{0}^{E_\mu} \!\!\!\rmd u\, x(u)\;I_0(\ell\sqrt{u}).\label{eq:JT_disk_function}
\end{align}
\subsection{Critical hyperbolic surfaces and the hyperbolic $O(n)$ loop model}\label{subsection:hyp_o(n)}
\begin{figure}[htbp]
\centering
\includegraphics[width=\textwidth]{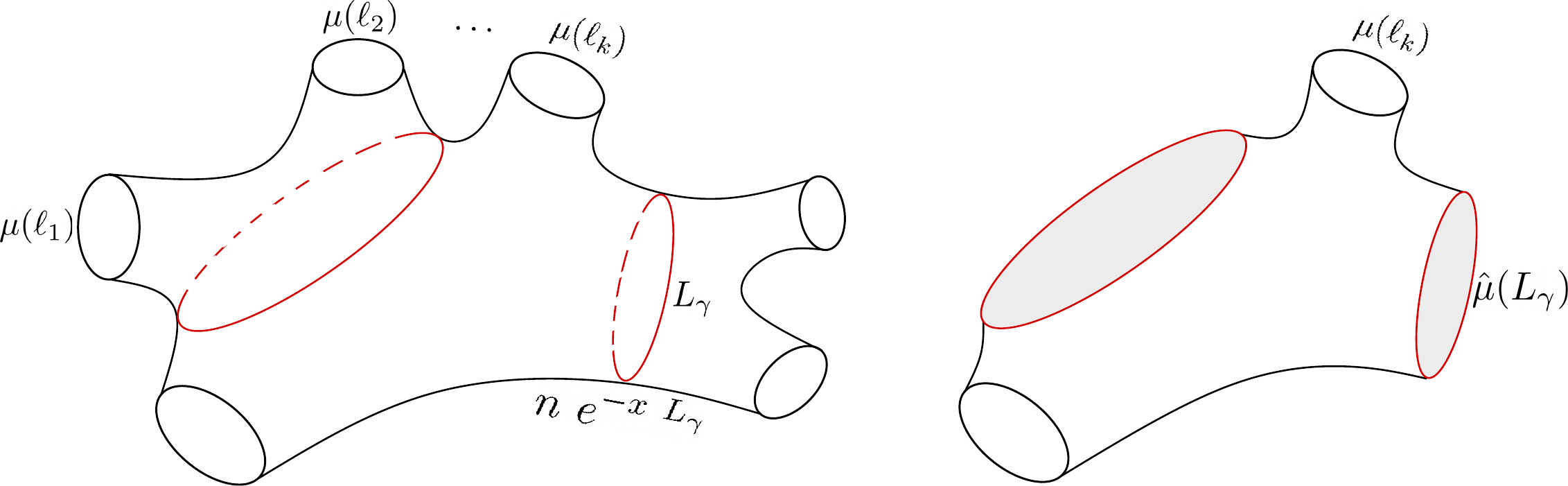}
\caption{\textbf{Left}: Example of a hyperbolic $O(n)$ loop model with two non-intersecting geodesic loops in red. \textbf{Right}: The corresponding gasket where the marked geodesic boundaries are colored in gray. \label{fig:hyper_o(n)}}
\end{figure}

\begin{figure}[htbp]
\centering
\includegraphics[width=.8\textwidth]{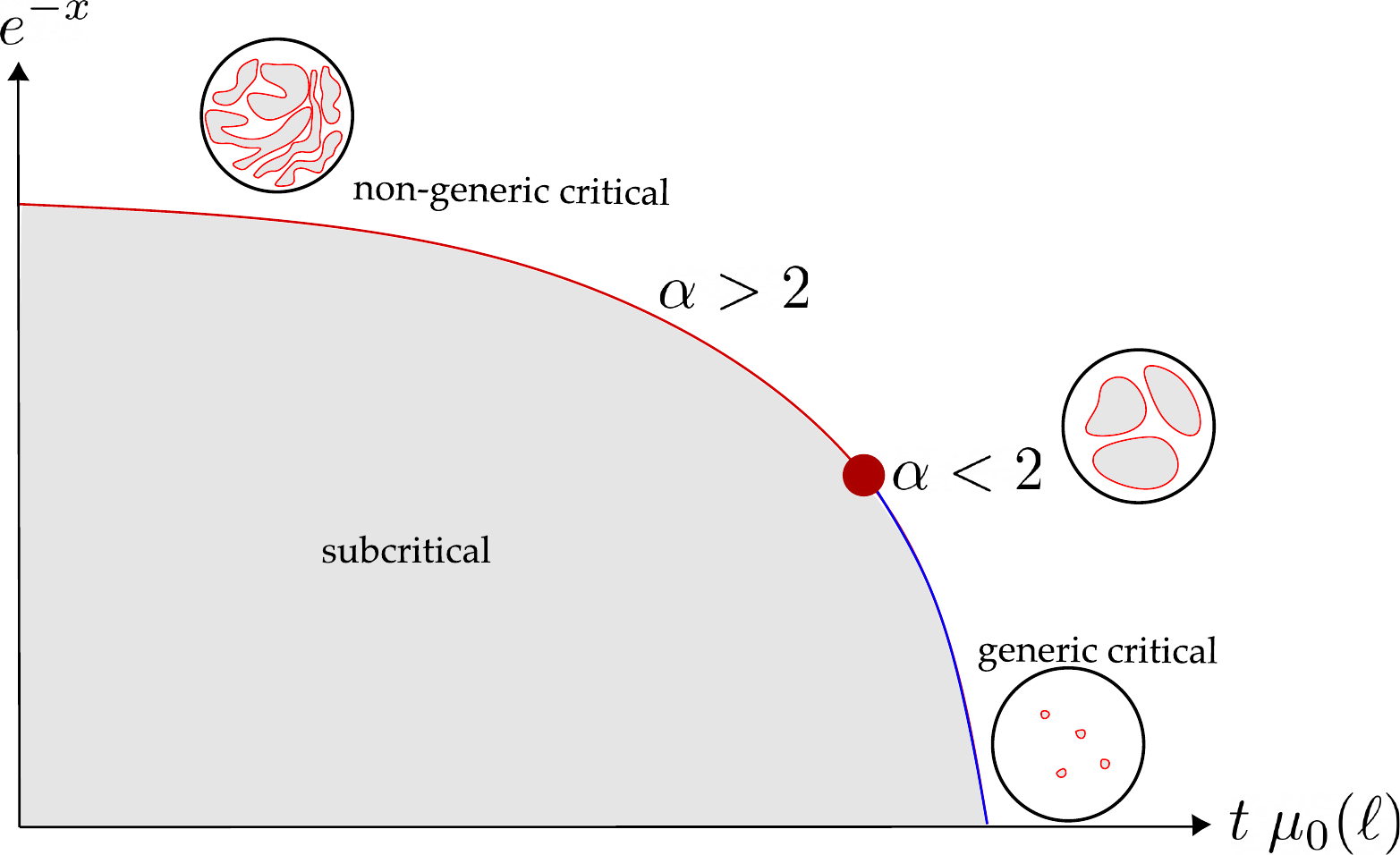}
\caption{Qualitative phase diagram of the $O(n)$ model \cite{Borot:2011kg,Borot_2012}. The line $x=2\sqrt{E_\mu}$ corresponds to non-generic criticality, $t>0$ and $\mu_0(\ell)$ is a reference weight. \label{fig:phase_diagram}}
\end{figure}

A natural question that arises is if a critical behavior, analogous to that observed in planar maps, happens for hyperbolic surfaces too. Specifically, whether there exist weights $\mu$ such that the string equation \eqref{eq:general_string_eq_JT} behaves in a similar manner to the ones presented in Table \ref{tab:critical}. The key difference between the Boltzmann maps and hyperbolic surfaces string equations is the background times $\gamma_k$. This implies that, in order to reach a non-generic critical regime in the hyperbolic case, the weight $\mu$ needs to be fine-tuned to dominate the Bessel function in the first term of \eqref{eq:general_string_eq_JT}, instead of the simpler linear term in the Boltzmann map case \eqref{eq:string_eq_maps}. Therefore, it appears to be a significantly larger challenge to find non-generic critical weights for hyperbolic surfaces. The existence of such regimes for hyperbolic surfaces is formally proved in \cite{budd2023}.  In this subsection, we will present an alternative argument for their existence, drawing upon an analogy with the gasket decomposition of the hyperbolic $O(n)$ loop model.\\
Consider the string equation \eqref{eq:general_string_eq_JT}, which, for convenience, we write again 
\begin{equation*}
 x(u)=\frac{\sqrt{u}}{\pi}J_1\left(2\pi\sqrt{u}\right)-\int_0^\infty \rmd\ell \mu(\ell) I_0\left(\ell \sqrt{u}\right),    
\end{equation*}
and a weight $\mu(\ell) = \kappa \,e^{-\lambda\ell} \hat{\mu}(\ell)$ with $\kappa,\lambda>0$ and $\hat{\mu}(\ell)$ a bounded, non-negative function on $[0,\infty)$ with asymptotics $\hat{\mu}(\ell) \sim C \,\ell^{-\alpha}$ as $\ell\to\infty$, e.g.\ $\hat{\mu}(\ell) = (\ell+1)^{-\alpha}$ or $\hat{\mu}(\ell) = \ell^{-\alpha} \Theta(\ell-1)$. We will demonstrate how we can choose constants $\kappa,\lambda>0$ to reach non-generic criticality, i.e.\ $x(E_\mu)=0$, $x'(E_\mu)=0$ and $x(u) \sim c\, (E_\mu-u)^{\alpha-1/2}$ as $u\to E_\mu$, with $\alpha \in (3/2,5/2)$.
To see this, we note that if we have $\mu$ such that $x'(E_\mu)=0$, then  $x''(E_\mu)=\infty$. This last expression requires that $\lambda = \sqrt{E_\mu}$.
Then, $x(E_\mu)=0$ allows us to fix $\kappa$ to
\begin{equation}
    \kappa = \frac{\frac{\sqrt{E_\mu}}{\pi} J_1(2\pi\sqrt{E_\mu})}{\int_0^\infty \rmd \ell \hat{\mu}(\ell) e^{-\ell\sqrt{E_\mu}}I_0(\ell \sqrt{E_\mu})}.\label{eq:kappa}
\end{equation}
If we take $E_\mu$ to be the first zero of
\begin{equation}
    x'(E_\mu) = J_0(2\pi\sqrt{E_\mu}) - \frac{\sqrt{E_\mu}}{\pi} J_1(2\pi\sqrt{E_\mu}) \frac{\int_0^\infty \rmd \ell \hat{\mu}(\ell) e^{-\ell\sqrt{E_\mu}}\frac{\ell I_1(\ell \sqrt{E_\mu})}{\sqrt{E_\mu}}}{\int_0^\infty \rmd \ell \hat{\mu}(\ell) e^{-\ell\sqrt{E_\mu}}I_0(\ell \sqrt{E_\mu})},
\end{equation}
which exists since $x'(0)=1$ and it is negative for $\sqrt{2}E_\mu = c_0^2/(8\pi^2) = 0.0732\ldots$, where $c_0$ is the first zero of the Bessel function $J_0$.
Accordingly,

\begin{align}
    x''(u) &= \left(\frac{\sqrt{u}}{\pi}J_1\left(2\pi\sqrt{u}\right)\right)'' - \int_0^\infty \rmd \ell \mu(\ell) \frac{\ell^2}{u}I_2(\ell\sqrt{u})\\
    &= (\cdots)''-\int_0^\infty \rmd \ell \ \kappa \ C \,e^{-\sqrt{E_\mu}\ell} \ell^{-\alpha} \frac{\ell^2}{u}\frac{e^{\sqrt{u}\ell}}{\sqrt{2\pi\sqrt{u}\ell}}+\mathcal{O}(\ell^{-\alpha+5/2})\\
    &= (\cdots)''-\frac{\kappa \ C}{u\sqrt{2\pi\sqrt{u}}} \Gamma\left(\frac{5}{2}-\alpha\right)\ (\sqrt{E_\mu}-\sqrt{u})^{\alpha - 5/2} + \mathcal{O}((\sqrt{E_\mu}-\sqrt{u})^{\alpha - 5/2})\\
    &= -\frac{\kappa \ C}{E_\mu \sqrt{2\pi\sqrt{E_\mu}}} \Gamma\left(\frac{5}{2}-\alpha\right)\ (2\sqrt{E_\mu})^{5/2-\alpha}(E_\mu-u)^{\alpha - 5/2} + \mathcal{O}((E_\mu-u)^{\alpha - 5/2}),
\end{align}
where we use Karamata's Tauberian Theorem (see e.g. \cite{feller-vol-2}, Section XIII.5, Theorem 2) in the last line. Integrating twice and using that $x(E_\mu)=x'(E_\mu)=0$ gives

\begin{align}
    x(u)&= -\frac{\kappa \ C}{\sqrt{2\pi}}\Gamma\left(\frac{1}{2}-\alpha\right) 2^{5/2-\alpha} E_\mu^{-\alpha/2} (E_\mu-u)^{\alpha-1/2}+\mathcal{O}((E_\mu-u)^{\alpha - 1/2})\\
    &=-\frac{\kappa \ C}{\sqrt{2\pi\sqrt{E_\mu}}}\Gamma\left(\frac{1}{2}-\alpha\right) 2^{2} (\sqrt{E_\mu}-\sqrt{u})^{\alpha-1/2}+\mathcal{O}((E_\mu-u)^{\alpha - 1/2})
\end{align}
This ends the demonstration. A similar reasoning follows for the disk function \eqref{eq:JT_disk_function}, 
\begin{align}
    W_\mu(\ell) &= -\frac{1}{2}\int_0^{E_\mu} \mathrm{d}u \,x(u) I_0(\ell\sqrt{u})\\
    &= \frac{1}{2} \frac{\kappa \ C}{\sqrt{2\pi\sqrt{E_\mu}}}\Gamma\left(\frac{1}{2}-\alpha\right) 2^{2} \int_0^{E_\mu} (\sqrt{E_\mu}-\sqrt{u})^{\alpha-1/2} \frac{e^{\ell \sqrt{u}}}{\sqrt{2\pi \ell\sqrt{u}}} \mathrm{d}u+ \cdots\\
    &=  \frac{\kappa \ C}{\sqrt{2\pi\sqrt{E_\mu}}}\Gamma\left(\frac{1}{2}-\alpha\right) 2^{2} \int_0^{\sqrt{E_\mu}} y^{\alpha-1/2} \frac{e^{-\ell y +\ell \sqrt{E_\mu}}}{\sqrt{2\pi \ell}}E_\mu^{1/4}\mathrm{d}y+ \cdots\\
    &= \frac{2}{\pi}\kappa \, C \, \Gamma\left(\frac{1}{2}-\alpha\right) \Gamma(\alpha + 1/2) \ell^{-\alpha-1}e^{\ell \sqrt{E_\mu}} + \mathcal{O}(\ell^{-\alpha-1})\\
     &= \frac{2}{\sin\left(\pi\left(\alpha-\frac{1}{2}\right)\right)}\kappa \, C \,  \ell^{-\alpha-1}e^{\ell \sqrt{E_\mu}} + \mathcal{O}(\ell^{-\alpha-1})\label{eq:disk_non-gen-crit}
\end{align}
where we used Euler's reflection formula for the gamma function to write the second line.
Now, we would like to show how this fits into the hyperbolic $O(n)$ loop model in a similar way as for Boltzmann planar maps \cite{Borot:2011kg}. The hyperbolic $O(n)$ loop model is defined by the generating function 
\begin{align}
    F^{(n)}_0[\mu] &= \sum_{k=3}^\infty\frac{1}{k!} \int_{0}^\infty \rmd \ell_1 \mu(\ell_1) \cdots \int_{0}^\infty \rmd \ell_k \mu(\ell_k)
	\sum_\Gamma \int_{\mathcal{M}_{0,k}(\ell_1,...,\ell_k)}\rmd_{WP} \prod_{\gamma\in \Gamma} n \ e^{-x L_\gamma}
\end{align}
where $n\in(0,2)$, $\Gamma$ is a collection of non-intersecting simple geodesic loops $\gamma$ of length $L_\gamma$ and $\rmd_{WP}$ is the Weil-Petersson measure of the hyperbolic surface. The coupling constant $n$ introduces a weight $n\ e^{-x L_\gamma}$ to each loop drawn on top of a hyperbolic surface (see Figure \ref{fig:hyper_o(n)}).\\  
What we aim to investigate is if there is a consistent weight substitution $\mu \rightarrow\hat{\mu}$ that relates the disk functions of the hyperbolic $O(n)$ loop model and the hyperbolic non-generic critical surfaces one, $W_\mu^{(n)}(\ell)=W_{\hat{\mu}}(\ell)$. From \eqref{eq:disk_non-gen-crit}, we can observe that our non-generic critical weight behaves asymptotically as
\begin{equation}
    \hat{\mu}(\ell)= 2\;\sin\left(\pi\left(\alpha-\frac{1}{2}\right)\right)  \ e^{-2\sqrt{E_\mu}\ell} \ell\; 2^{-2}\; W_{\hat{\mu}}(\ell)  \label{weigth_substitution_hypo(n)}
\end{equation}
Therefore, we have proven that a weight substitution 
\begin{equation}
    \hat{\mu}(L_\gamma) = \mu(L_\gamma) + n  \ e^{-L_\gamma x} L_\gamma \; 2^{-2}\;W_{\hat{\mu}}(L_\gamma),\label{eq:on_non_gen_critical}
\end{equation}
for $x=2\sqrt{E_\mu}$ and $n=2\sin\left(\pi\left(\alpha-\frac{1}{2}\right)\right)$\footnote{The factor of $2^{-2}$ is necessary for consistency with the prefactor in the generating function of Weil-Petersson volumes and in \eqref{eq:JT_disk_function}. Thus, it does not affect the relation between $n$ and $\alpha$.} consistent with the disk function substitution and with asymptotics $\hat{\mu}(L_\gamma)\sim C' e^{\sqrt{E_\mu }L_\gamma}L_\gamma^{-\alpha}$ exists. In geometrical terms, this means that one can substitute each part of the hyperbolic surface surrounded by a geodesic loop $\gamma$ by a geodesic boundary with weight \eqref{eq:on_non_gen_critical} (see Figure \ref{fig:holes_maps}).\\
\\
To conclude this chapter, we present the qualitative phase diagram of the $O(n)$ loop model (Figure \ref{fig:phase_diagram}) as studied in \cite{Borot:2011kg, Borot_2012}, which we expect to generalise for the hyperbolic $O(n)$ loop model. This will be relevant when we identify non-generic criticality in JT gravity (Section \ref{Section_Criticality}) and draw  similarities with hyperbolic geometries. \\
As we can see in Figure, \ref{fig:phase_diagram}, within the non-generic critical family of weights, there are two cases, the \textbf{\textit{dense}} and \textbf{\textit{dilute}} phases. In the dense phase ($\alpha>2$), the length of the loops is big and they can touch tangentially. In the dilute phase ($\alpha<2$), the loops tend to avoid each other and also themselves. 
\section{JT Gravity}\label{section_JT_random}
Let us turn now to the standard formulation of JT gravity before reintroducing it in terms of random geometry. JT gravity is a 2-dimensional theory of gravity defined by the action 
\begin{equation}
    I_{JT}=-\frac{S_0}{2\pi}\left(\frac{1}{2}\int_\mathcal{M}\sqrt{g}R+\int_{\partial\mathcal{M}}\sqrt{h}K\right)-\left(\frac{1}{2}\int_\mathcal{M}\sqrt{g} \phi (R+2) +\int_{\partial\mathcal{M}} \sqrt{h} \phi (K-1)\right)\label{eq:JTaction},
\end{equation}
where the first term is the Einstein-Hilbert action, which is topological in 2D and is proportional to the Euler characteristic of the surface $\mathcal{M}$, $\chi(\mathcal{M})=2-2g-n$, where $g$ is the genus of the surface and $n$ is the number of boundaries. The second term contains a scalar field $\phi$, called \textit{dilaton}, which acts as a Lagrange multiplier that imposes a negative curvature $R=-2$ in the bulk, this is called AdS spacetime. This is equivalent to imposing $\mathcal{M}$ to be a hyperbolic surface of genus $g$ and a boundary $\partial \mathcal{M}$ composed by a disjoint union of $n$ asymptotic boundaries with extrinsic curvature $K$ \cite{Saad:2019lba}. It is worth noting that the term ``asymptotic boundary" accounts for the fact that AdS spacetime is infinite in extent and does not have a traditional boundary that restricts or limits the spacetime itself but has a conformal boundary located at infinity and serves as a reference for defining the conformal structure of the spacetime.\\
Hence, the $n$-boundary JT gravity partition function reduces to a sum over geometries with asymptotic boundaries of length $(\beta_1,\dots, \beta_n)$ which, after integrating out the dilaton in the bulk, is
\begin{equation}
    Z(\beta_1,\dots,\beta_n)=\int_\mathcal{M} \mathcal{D}g\hspace{1mm}\mathcal{D}\phi \hspace{1mm} e^{-I_{JT}}= \int_{\mathcal{M}}
  \mathcal{D}g \hspace{1mm} e^{S_0 \chi(\mathcal{M})} \delta(R+2) \int\mathcal{D}\phi \hspace{1mm} e^{\int_{\partial\mathcal{M}} \sqrt{h} \phi (K-1)}.\label{eq:JTpathintegral}
\end{equation}
The last expression hints at a topological expansion of spacetimes that split between bulk and boundary dynamics. It geometrically translates to the existence of a geodesic closed curve of length $b_i$ per asymptotic boundary $\beta_i$ that split a hyperbolic surface into two parts: bulk and \textit{trumpets}. The bulk part of \eqref{eq:JTpathintegral} is an integral over the moduli space of hyperbolic surfaces of genus $g$ and $n$ geodesic boundaries, and it is given by the Weil-Petersson volume $V^{WP}_{g,n}(b_1,\dots,b_n)$ \cite{Mirzakhani}. 
The trumpet is the part of the geometry that connects the geodesic boundaries with the asymptotic boundaries. The asymptotic boundary term in \eqref{eq:JTpathintegral} is given by the Schwarzian derivative of the near-boundary coordinates, which encodes information about the character of the asymptotic conformal symmetry group. Additionally, from the holographic point of view, it effectively describes a one-dimensional quantum mechanical system given by the so-called Schwarzian action. Then, the trumpet partition function is the partition function for a geometry with the Schwarzian action in the asymptotic boundary ending on a geodesic boundary, and it is given by
\begin{equation}
    Z_{tr}(\beta,b)=\frac{e^{-b^2/4\beta}}{2\sqrt{\pi \beta}}.
\end{equation}
Therefore, \eqref{eq:JTpathintegral} can be written as 
\begin{equation}
    Z(\beta_1,\dots,\beta_n)=\sum_{g=0}^\infty e^{S_0 \chi}\int_0^\infty b_1 \rmd b_1 \dots \int_0^\infty b_n\rmd b_n V_{g,n}(b_1,\dots,b_n)Z_{tr}(\beta_1,b_1)\dots Z_{tr}(\beta_n,b_n).\label{eq:JTpart_g_n}
\end{equation}
From now on we will focus on the case $n=1$, i.e.\ one asymptotic boundary. In this case, \eqref{eq:JTpart_g_n} becomes 
\begin{equation}
    Z(\beta)=\sum_{g=0}^\infty e^{-S_0(2g-1)} \tilde{Z}_g(\beta)= e^{S_0} \tilde{Z}_0(\beta) + \sum_{g=1}^\infty e^{-S_0(2g-1)}\int_0^\infty b\rmd b V^{WP}_{g,1}(b)Z_{tr}(\beta,b).
\end{equation}
We separate the contribution for $g=0$ because $V_{0,1}$ vanishes. However, this geometry corresponds to a hyperbolic disk for which the path integral \eqref{eq:JTpathintegral} is not vanishing, this is what we call $\tilde{Z}_0(\beta)$ in this case, and it is given by \cite{Saad:2019lba}
\begin{equation}
    \tilde{Z}_0(\beta)=\frac{1}{\sqrt{16\pi\beta^3}}e^{\frac{\pi^2}{\beta}}
\end{equation}
\subsection{JT Gravity with defects}
\begin{figure}[htbp]
\centering
\includegraphics[width=.9\textwidth]{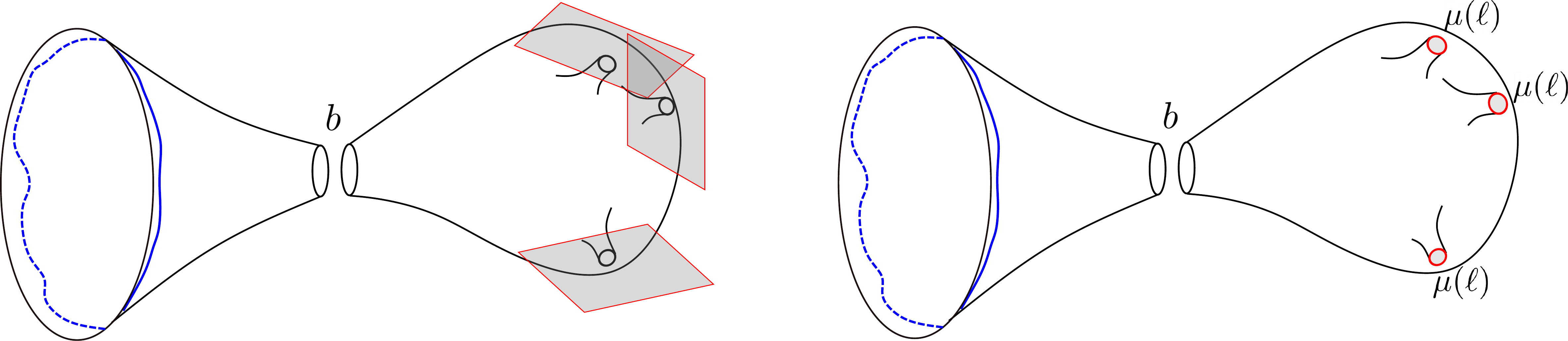}
\caption{Left: JT spacetime ending on $n$ branes. Right: JT spacetime with $n$ geodesic boundaries with weight $\mu$. \label{fig:branes}}
\end{figure}

\begin{figure}[htbp]
\centering
\includegraphics[width=\textwidth]{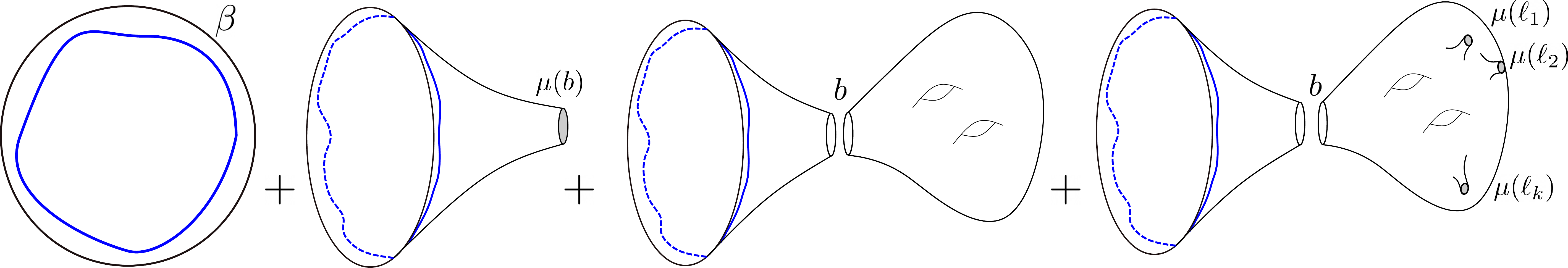}
\caption{Examples of geometries that contribute to the partition function \eqref{eq:full_JT_partfunction}. From left to right: $(g=0,n=1,k=0)$, $(g=0,n=1,k=1)$, $(g=2,n=1,k=0)$ and $(g=2,n=1,k>0)$. \label{fig:JT_full_sum}}
\end{figure}

In a similar fashion, one can consider the case when the hyperbolic surface has geodesic boundaries of length $\ell$ that are not attached to an asymptotic boundary by a trumpet, these are called \textit{defects} \cite{Turi:2020ale} and are assigned an effective weight $\mu(\ell)$ that compile the physical system they represent e.g. conical defects or branes (See Fig. \ref{fig:branes}). 
Then, the partition function can be expanded in a sum over genus and a sum over the number of defects 
\begin{align}
    Z(\beta) =\sum_{g=0}^\infty e^{-S_0(2g-1)}\sum_{k=0}^{\infty} e^{-S_0 k} Z_{g,k+1}(\beta).\label{eq:full_JT_partfunction}
\end{align}
where for $g>0$ and $k\geq 0$,
\begin{equation}
    Z_{g,k+1}(\beta)=\frac{1}{k!}\int_0^\infty \rmd \ell_1
\mu(\ell_1)...\int_0^\infty \rmd \ell_k
\mu(\ell_k)\int_0^\infty b\rmd b V^{WP}_{0,k+1}(b,\ell_1,...,\ell_k)Z_{tr} (\beta,b)
\end{equation}
In particular, for $g=0$ we know that the Weil-Petersson volumes vanish for the cases $k=0,1$, but these partition functions can be explicitly computed using the Schwarzian action and they correspond to the disk ($\tilde{Z}_0$) and the trumpet ending on a defect (sometimes called \textit{half-wormhole}). Therefore, the genus-$0$ partition function can be explicitly written as 
\begin{align}
    Z_0(\beta)
&=\frac{1}{\sqrt{16\pi\beta^3}}e^{\frac{\pi^2}{\beta}}+e^{-S_0}\int_0^\infty \rmd b\mu(b)\frac{e^{-b^2/4\beta}}{2\sqrt{\pi \beta}}\\ 
    &+ \sum_{k=2}^\infty \frac{e^{-S_0k}}{k!}\int_0^\infty \rmd \ell_1
\mu(\ell_1)...\int_0^\infty \rmd \ell_k
\mu(\ell_k)\int_0^\infty b\rmd b\frac{e^{-b^2/4\beta}}{2\sqrt{\pi \beta}}V^{WP}_{0,k+1}(b,\ell_1,...,\ell_k).\label{eq:full_diskamplitude}
\end{align}
More examples of the geometries that contribute to \eqref{eq:full_JT_partfunction} are shown in Figure \ref{fig:JT_full_sum}.\\
\\
So far, we have introduced JT gravity as it is mainly done in the physics literature, yet, one of our main goals is to reintroduce techniques from random geometry that we will prove are highly useful for JT gravity (Section \ref{Section_Criticality}). In particular, the power of the string equation formulation has been used in JT gravity \cite{Turiaci:2020fjj,Forste:2021roo}\footnote{It is worth noting that what we call the string equation is known in the physics literature as the genus $g=0$ string equation in the limit $\hbar\rightarrow 0$}, but, in the physics literature, one usually computes the partition function and density of states using a holographic quantum mechanical system in the asymptotic boundary and then `guess' the form of the string equation. Here, we take the opposite point of view and start with the string equation.\\
Using this, we can re-express \eqref{eq:full_diskamplitude} in a more convenient way which explicitly makes use of the string equation 
\cite{Johnson:2019eik,Okuyama:2019xbv,Turi:2020ale}
\begin{align}
 Z(\beta)_{g=0}&=\frac{1}{2\sqrt{\pi\beta}}\int_{-\infty}^0\rmd x e^{u(x) \beta }=
\frac{1}{2\sqrt{\pi\beta}}\int_{-\infty}^{E_\mu}\rmd u e^{u \beta }\frac{\partial x(u)}{\partial u}\label{eq:Z0def}\\
&=  -\frac{1}{2}\int_{-\infty}^0\rmd u e^{u \beta }x(u)\sqrt{\frac{\beta}{\pi}} -\frac{1}{2}\int_{0}^{E_\mu}\rmd u e^{u \beta }x(u)\sqrt{\frac{\beta}{\pi}}\\
&=  \frac{1}{\sqrt{16\pi\beta^3}}e^{\frac{\pi^2}{\beta}}+\int_0^\infty \rmd b\mu(b)\frac{e^{-b^2/4\beta}}{2\sqrt{\pi \beta}} + \int_0^\infty b\rmd b\frac{e^{-b^2/4\beta}}{2\sqrt{\pi \beta}}W(b),
\end{align}
where we set $e^{S_0}=1$ for convenience and in the last line we used the identity
\begin{equation}
    e^{-bz}=2z\int_0^\infty \rmd \beta e^{-\beta z^2}\frac{e^{-b^2/4\beta}}{2\sqrt{\pi \beta}}.
\end{equation}
The density of states is obtained by Laplace transforming the partition function, and is given by
\begin{equation}
        \rho_{0}(E) = \frac{1}{2\pi}\int_{-E}^{-E_0}\rmd u \frac{1}{\sqrt{E+u}} \frac{\partial x}{\partial u}\label{eq:rho0def}.
\end{equation}
In order to draw a parallel between this language and some well-known results in the JT gravity literature, we present two examples: FZZT and end-of-the-world branes.
\subsection{Example: FZZT Branes}
Dynamical FZZT Branes were introduced in the context of Liouville Field Theory in \cite{Fateev:2000ik} as a family of conformally invariant (Dirichlet) boundary conditions parametrized by the so-called \textit{boundary cosmological constant}, which controls the length of the boundary. Beyond the Physics aspects, the convenience of working with these branes comes from the fact that they have well-defined analytic expressions in Matrix Models, they are determinant insertions \cite{Aganagic:2003qj}. In the JT gravity literature, FZZT branes have been studied in this language \cite{Saad:2019lba,Blommaert:2019wfy}, and it was shown that these branes can be effectively encoded by geodesic boundaries of length $\ell$ with weight \cite{Okuyama:2021FZZT} (see Figure \ref{fig:branes})
\begin{equation}
    \mu_{FZZT}(\ell)=-e^{-z \ell}.
\end{equation}
where the constant $z$ plays the role of the boundary cosmological constant. Using \eqref{eq:def_general_JT_times}, we can compute the times 
\begin{equation}
    t_k^{FZZT}=-\frac{4^{-k}(2k)!}{k!}z^{-2 k-1}, 
\end{equation}
which are consistent with \cite{Okuyama:2021FZZT}. Then, the string equation \eqref{eq:general_string_eq_JT} takes the well-known form
\begin{equation}
     x(u)=\frac{\sqrt{u}}{\pi}J_1\left(2\pi\sqrt{u}\right)+\frac{1}{\sqrt{z^2-u}}.
\end{equation}
It is worth noting that the term `dynamical' in this context means that all physical quantities, e.g. \eqref{eq:Z0def} and \eqref{eq:rho0def}, are computed using the shifted weights $t_m+\gamma_m$ instead of the pure JT gravity ones, $\gamma_m$. In these terms is that our computations are also done in the dynamical framework.\\
\subsection{Example: EOW branes}
In JT gravity, End-of-the-world branes are boundaries of space where Neumann boundary conditions are usually imposed. These have been mainly used to investigate black hole evaporation and the Page curve\cite{Penington:2019kki}, and correspond to introducing vectors in the JT gravity Matrix Model. However, for our purposes, EOW branes are geodesic boundaries with weights \cite{Gao:2021uro}
\begin{equation}
    \mu_
{EOW}(\ell)=\frac{e^{-m\ell}}{2\sinh{\left(\ell/2\right)}}\label{eq:eow_weight}
\end{equation}
and, equivalently, times 
\begin{equation}
    t^{EOW}_k=\frac{4^{-k}(2k)! \zeta \left(2 k+1,m+\frac{1}{2}\right)}{k!} \mathrm{1}_{k\geq 1},
\end{equation}
where $m$ is the value of the normal derivative of the dilaton at the EOW boundary.


\section{Critical JT Gravity}\label{Section_Criticality}
Our objective is to investigate the applicability of the criticality results observed in Boltzmann planar maps and more recently in hyperbolic surfaces \cite{budd2023} to JT gravity. In this last work, it is proven that the generating function of Weil-Petersson volumes experiences a similar critical behavior to Boltzmann planar maps (see Table \ref{tab:critical}), a result that we also showed in Subsection \ref{subsection:hyp_o(n)} at the level of the string equation and disk function. Furthermore, in the non-generic critical regime ($M_0^{WP}=0$), the hyperbolic surfaces also develop large faces, which correspond to geodesic boundaries. We are especially interested in how the non-generic criticality in the bulk translates into the holographic quantum system on the asymptotic boundary, and, in particular, how it affects its partition function and density of states. More precisely, we will show that this family of non-generic critical models consistently interpolates between two known cases in the JT literature, $\rho(E)\sim \sqrt{E}$ and $\rho(E)\sim E^{3/2}$, which correspond to the $k=1$ and $k=2$ minimal models respectively.\\
For simplicity, we will assume a general form of the non-generic critical weight $\mu_{\alpha}$ that satisfies the conditions deduced in Section \ref{section:Random_geometry} such that
the string equation is  
\begin{equation}
    x_{crit}(u)= -t_{crit}\hspace{1mm}(E_\mu-u)^{\alpha -\frac{1}{2}} + \mathcal{O}((E_\mu-u)^{\alpha -\frac{1}{2}})\label{muform}
\end{equation}
where $3/2<\alpha<5/2$. 
The genus-$0$ generating function is
\begin{equation}
    F^{WP}_0[\mu_\alpha]=\frac{E_\mu^{2 \alpha }}{8 \alpha } t_{crit} +\mathcal{O}(E_\mu^{2\alpha}),
\end{equation}
and the disk function 
\begin{equation}
    W(b)=\frac{E_\mu^{\alpha +\frac{1}{2}} \, _0F_1\left(\alpha +\frac{3}{2};\frac{b^2 E_\mu}{4}\right)}{2 \alpha +1} t_{crit} + \mathcal{O}(b^{-\alpha-1}).
\end{equation}
\begin{figure}[htbp]
\centering
\includegraphics[width=0.8\textwidth]{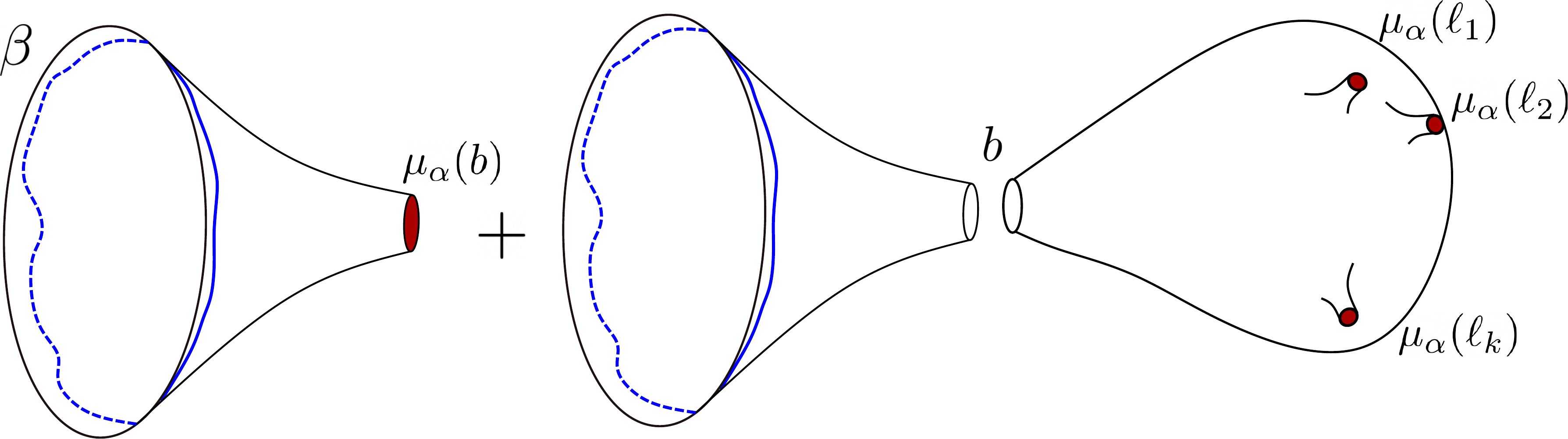}
\caption{Geometries that contribute to the partition function of critical JT gravity with defects of weight $\mu_\alpha$.\label{fig:criticalJT}}
\end{figure}\\

\subsection{Density of states}
\begin{figure}[htbp]
\centering
\includegraphics[width=.8\textwidth]{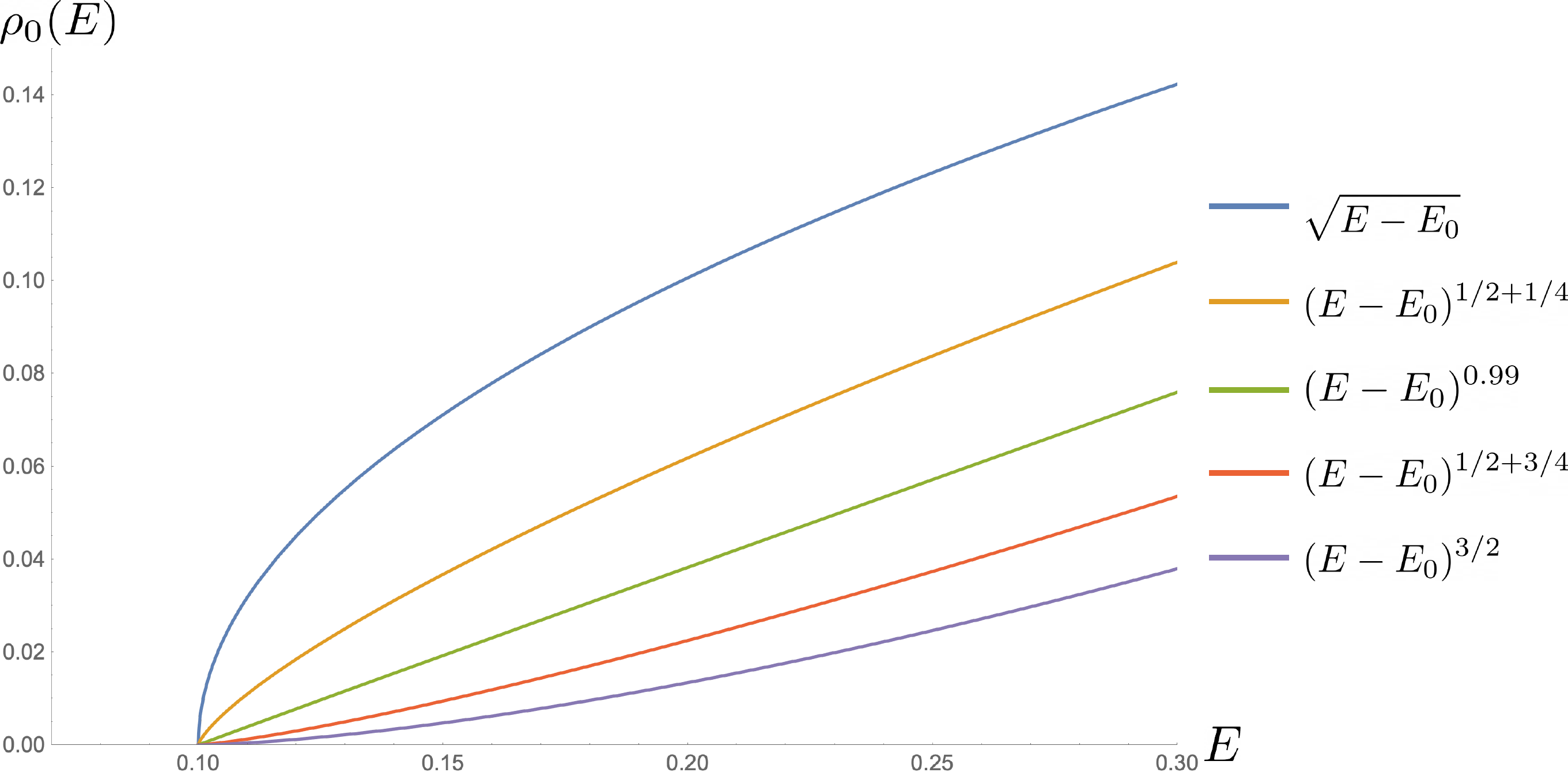}
\caption{Genus zero density of states \eqref{eq:rho_E} for different values of $\alpha$ and $t_{crit}=1$. This family of non-generic critical models continuously interpolates between $\sqrt{E-E_0}$ and $(E-E_0)^{3/2}$. Our simplified forms of the weight and string equation do not fix a value for $E_0$. For convenience, we consider the threshold energy as constant ($E_0=0.1$) for all $\alpha$, but, in general, it will also be a function of $\alpha$.  \label{fig:critical_density}}
\end{figure}

As emphasized in Section \ref{section_JT_random}, the string equation provides sufficient information for determining the leading-order genus zero partition function and density of states. So, using \eqref{muform}, \eqref{eq:Z0def} and \eqref{eq:rho0def}, and neglecting terms of higher order, the partition function of critical JT gravity is 
\begin{equation}
    Z_0(\beta)= \frac{ \Gamma \left(\alpha +\frac{1}{2}\right) e^{-\beta  E_0}t_{crit}}{2 \sqrt{\pi }}\beta ^{-\alpha }\label{eq:Z_beta},
\end{equation}
and its density of states is given by \eqref{eq:rho0def}, 
\begin{equation}
     \rho_{0}(E) = \frac{\Gamma \left(\alpha +\frac{1}{2}\right) t_{crit}}{2 \sqrt{\pi } \Gamma (\alpha )}(E-E_0)^{\alpha -1}\label{eq:rho_E},
\end{equation}
where we substituted $E_\mu\rightarrow -E_0$. In Figure \ref{fig:critical_density}, we show \eqref{eq:rho_E} for different values of $\alpha$. We note that this non-generic critical family of models interpolates constantly between $\sqrt{E-E_0}$ and $(E-E_0)^{3/2}$.\\

\subsection{Phase transitions}
\begin{figure}[htbp]
   \centering
   \includegraphics[width=.9\textwidth]{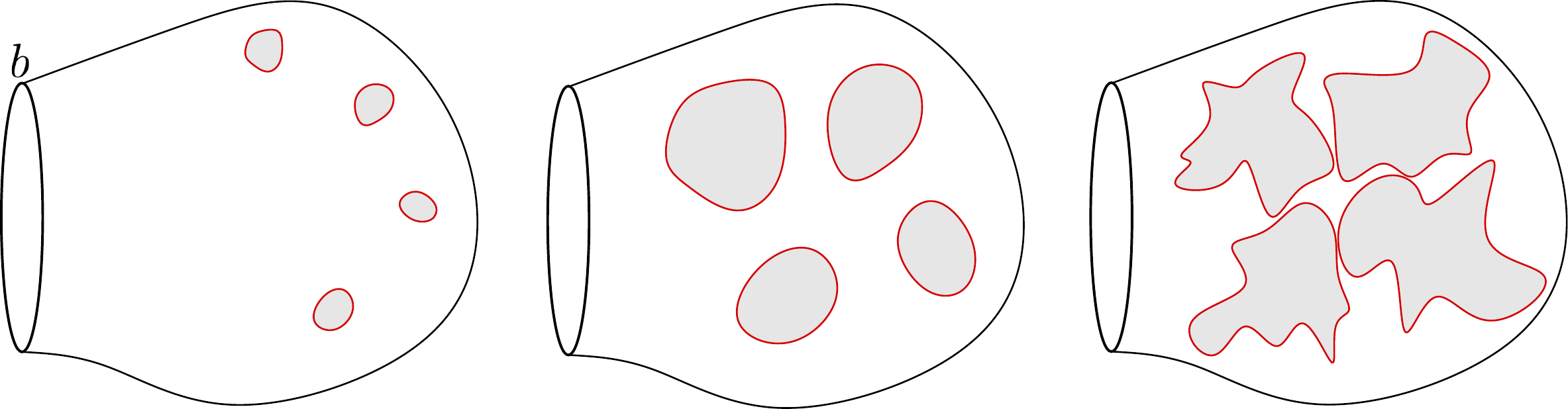}
   \caption{Schematic illustration of a hyperbolic surface with: subcritical, dilute, and dense non-generic critical weights.\label{fig:holes_JT}}
  \end{figure}

We dedicate this subsection to clarify the physical and geometrical meaning of the JT gravity non-generic critical family of models described by the density of states \eqref{eq:rho_E}. First, let us remark that it is known that in JT gravity with FZZT \cite{Okuyama:2021Page} and EOW \cite{Gao:2021uro} branes some parameters can be tuned to shift $\rho_0(E)$ from $\sqrt{E-E_0}$ to $(E-E_0)^{3/2}$, and our model bridges the gap between these two cases by providing an interpolation. However, the framework that we present in this paper makes clear and precise in which sense there is a phase transition. The explanation is that by fine-tuning the first derivative of the string equation to vanish, the weight goes from a subcritical to a generic critical phase. What our model does is fine-tune the weight to be non-generic critical. In principle, this non-generic critical weight can be linked to modifications to the FZZT and EOW brane weights. Given that our model offers this interpolation, it seems plausible this non-generic critical weight can be linked to deformations of the FZZT and EOW brane weights and its precise form could yield interesting results. However, we will not explore this direction in the present work, but rather the general properties of these deformations.\\
The aspect we highlight now is the geometrical features of this family of non-generic critical models. Within this critical regime, the system undergoes significant transformations. Notably, the analysis in \cite{budd2023} reveals formal divergences in the expected number of boundaries and their corresponding total length, indicating substantial geometrical similarities with the Boltzmann planar map case as the model approaches criticality.
We can see this by using the probability measure \eqref{eq:prob_density_WP} to compute the expected number of geodesic boundaries  
        \begin{equation}
           \mathbb{E}(n) = \sum_{n=3}^{\infty} n\cdot\frac{2^{2-n}}{n!} \frac{\prod_{i=1}^{n} \int_{0}^{\infty}\rmd \ell_i \mu(\ell_i) V^{WP}_{0,n}(\ell_1 , ... , \ell_n)}{F^{WP}_0},
        \end{equation}
The main argument is that when two of the defects have length zero, we can easily see that  
\begin{equation}
    \begin{split}
        \frac{\delta E_\mu}{\delta \mu(\ell)} =
        \sum_{n=3}^{\infty} \frac{2^{3-n}}{(n-3)!} \left( \prod_{k=4}^{n} \int_{0}^{\infty}  \rmd \ell_i \ \mu\left(\ell_i\right) \right) V^{WP}_{0,n}\left(0, 0, \ell, \ell_4, ..., \ell_{n}\right) \rightarrow \infty
    \end{split}
    \label{eq: R dmu}
\end{equation}
since, from the string equation, we have that $ \frac{\delta E_\mu}{\delta \mu(\ell)} = \frac{\delta u/\delta \mu (E_\mu)}{M^{WP}_0}$, and $M^{WP}_0=0$ at criticality. Therefore, $\mathbb{E}(n)\rightarrow\infty$.
Additionally, we can analyse the expected number of boundaries in a disk of diameter $b$ in the non-generic critical regime,
\begin{equation}
    \mathbb{E}_{W_\mu(b)}(n)=\int_0^\infty \rmd \ell \frac{\delta W_\mu(b)}{\delta \mu (\ell)}\mu(\ell) 	\propto \int_0^\infty \rmd \ell \; \ell^{-1/2}e^{\ell\sqrt{E_\mu}}\; \ell^{-\alpha}e^{-\ell \sqrt{E_\mu}} + \mathcal{O}(\ell^{-\alpha-1/2}).
\end{equation}
The expression inside the integral is the large $\ell$ probability distribution of the number of geodesic boundaries of size $\ell$ in a disk of diameter $b$. Then, we can see that the variance of this distribution is $\text{Var}(\ell^{-\alpha-1/2})\rightarrow\infty$ for $\alpha>3/2$. This implies that in the non-generic critical regime of JT gravity, one can expect a very large number of defects, and if $b\rightarrow\infty$ there will always be boundaries of comparable size. 
Even though we are not explicitly proving the existence of a dilute and dense phase, we anticipate a similar behavior to Boltzmann maps due to the relationship with the hyperbolic $O(n)$ loop model discussed in Section \ref{section:Random_geometry}. A schematic illustration of this behavior is provided in Figure \ref{fig:holes_JT}.\\






\section{Critical Dilaton Gravity}\label{Section_Dilaton_gravity}
\begin{figure}[htbp]
\centering
\includegraphics[width=\textwidth]{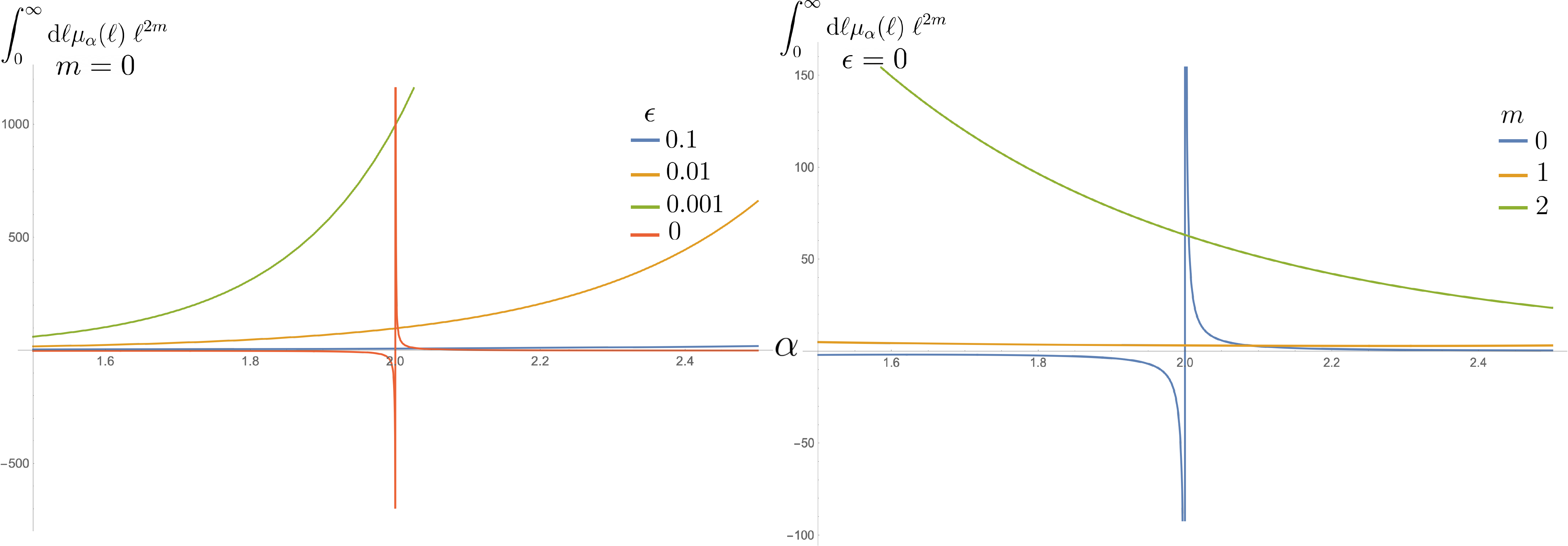}
\caption{Left: The $m=0$ coefficient \eqref{eq:dilaton_coeffs} for decreasing values of $\epsilon$. Right: First three coefficients of the potential \eqref{eq:dilaton_pot_series} for $\epsilon=0$ as functions of $\alpha$. In both cases, $E_\mu=0.1$ and $t_{crit}=1$.    \label{fig:dilaton_potential_coeff}}
\end{figure}

\begin{figure}[htbp]
\centering
\includegraphics[width=.9\textwidth]{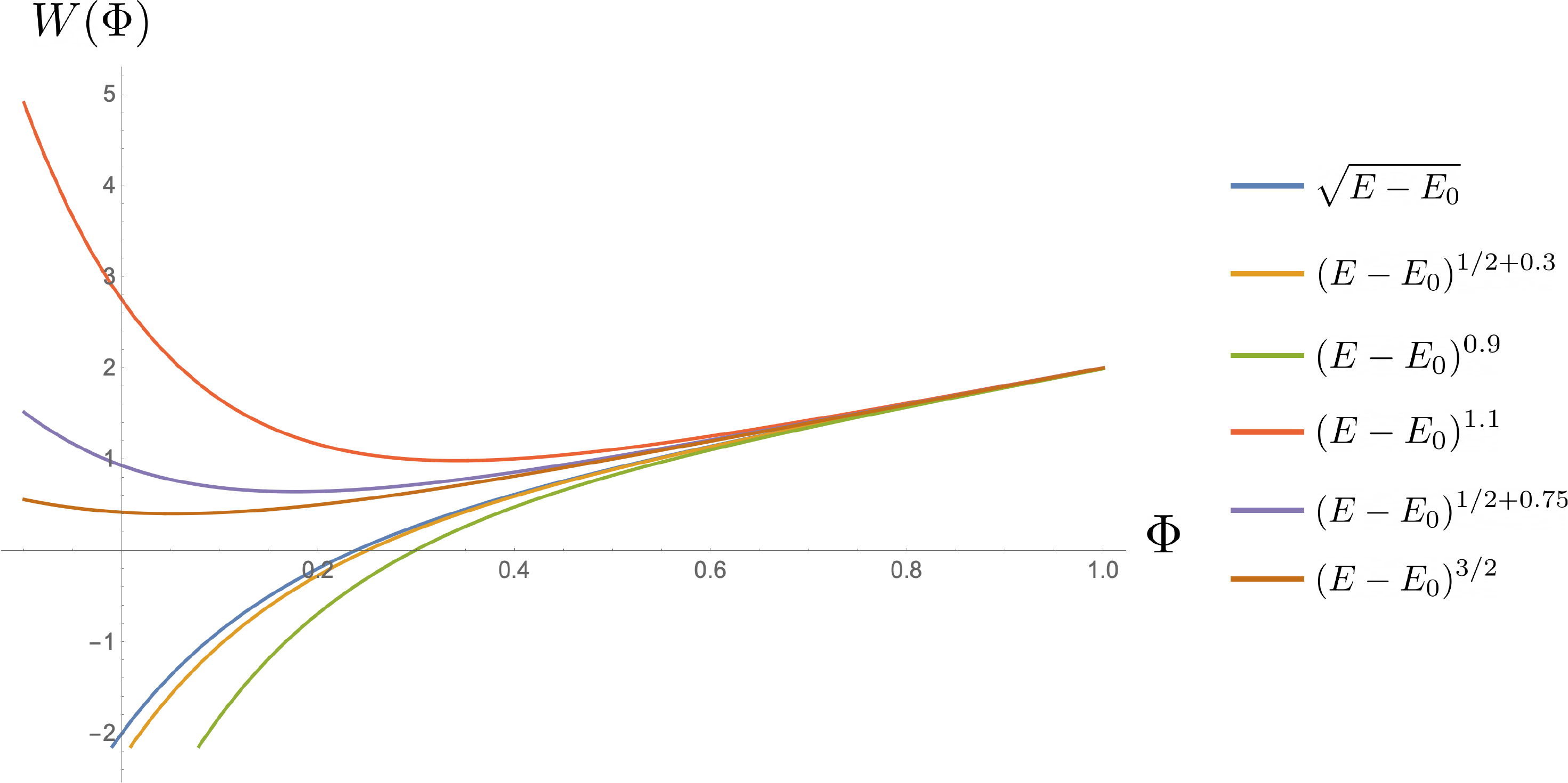}
\caption{Dilaton potential \eqref{eq:dilaton_pot_critical_particular} for different values of $\alpha$, $t_{crit}=1$ and $E_\mu=0.1$. We label the graphs using the corresponding energy density they correspond to.   \label{fig:dilaton_potential}}
\end{figure}

\begin{figure}[htbp]
\centering
\includegraphics[width=.8\textwidth]{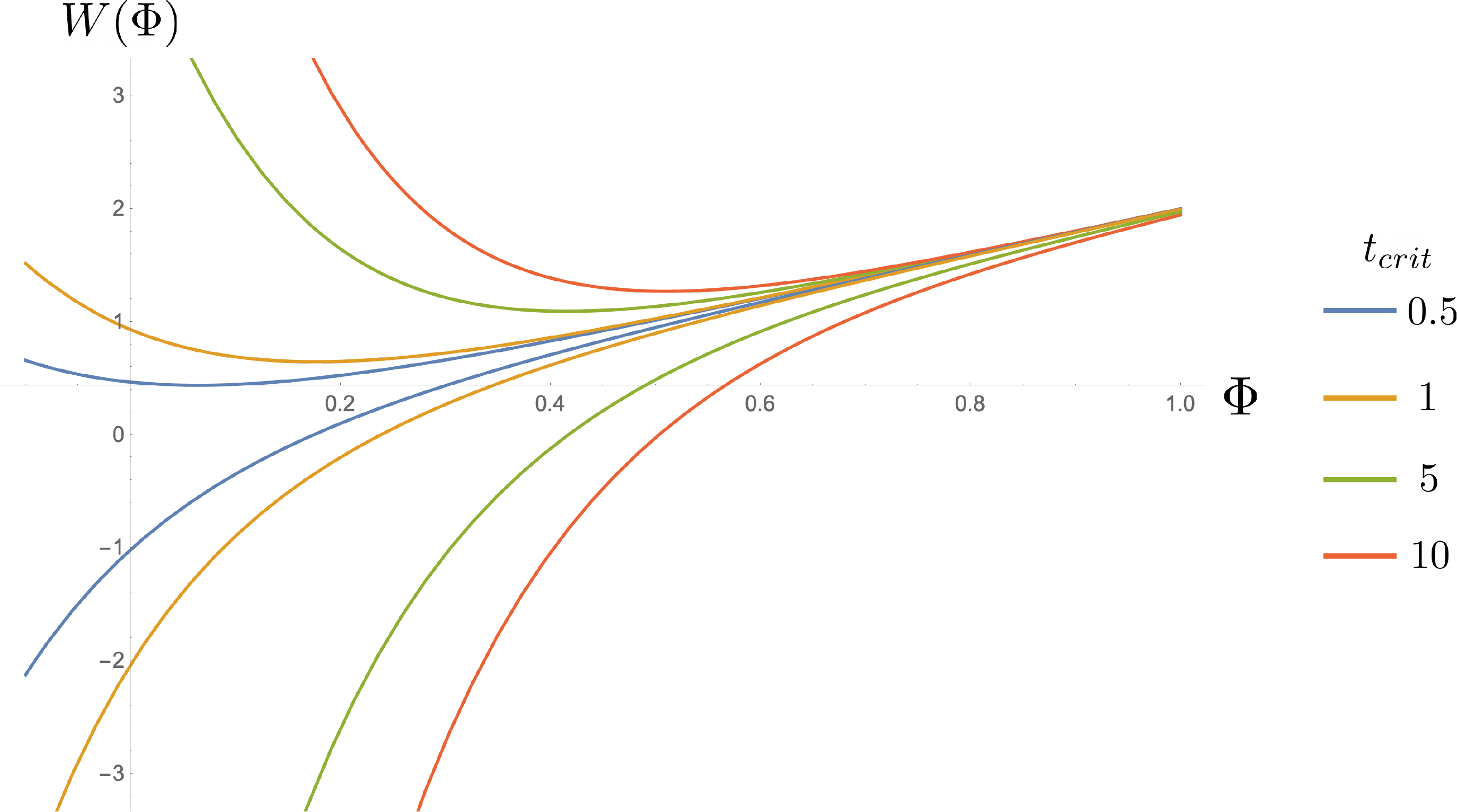}
\caption{Dilaton potential \eqref{eq:dilaton_pot_critical_particular} for different values of $t_{crit}$ and $E_\mu=0.1$. The top four graphs correspond to $\alpha=2.25$ ($(E-E_0)^{5/4}$)and the bottom four correspond to $\alpha=1.75$ ($(E-E_0)^{3/4}$).  \label{fig:dilaton_potential_t}}
\end{figure}

In the field of JT gravity with defects, such as conical defects and branes,  it has been proposed that these can be integrated out, leading to an effective dilaton action \cite{Witten:2020wvy, Turi:2020ale, Gao:2021uro}, which is a deformation of pure JT gravity. While the formality of this approach may remain subject to an ongoing debate, we adopt this particular conjectural view as a means to analyze non-generic critical models within the field of thermodynamic solutions in dilaton gravity.\\
Dilaton gravity is a 2-dimensional theory defined by the action 
\begin{equation}
    I_{dil}=\frac{1}{2}\int_{\mathcal{M}}\sqrt{g}(\Phi R+W(\Phi)).\label{eq:dilaton_action}
\end{equation}
where $\Phi$ is the dilaton and the potential is given by\footnote{\cite{Okuyama:2021FZZT} highlights the caveats of constructing the dilaton potential for a time modification different from the conical defect.}
    \begin{equation}    W(\Phi)=2\Phi+e^{-2\pi\Phi}\int_0^\infty \rmd \ell \mu(\ell)\cos\left(\ell \Phi\right).\label{eq:dilaton_pot_general}
    \end{equation}
The first term corresponds to the pure JT gravity case and the second term is determined by the weight per geodesic boundary. This potential satisfies the asymptotically AdS$_2$ condition, i.e.\ $W(\Phi)\sim 2\Phi$ as $\Phi\rightarrow +\infty$. Yet, its closed form  is only known for a few cases e.g. FZZT branes.\\
We can also write it as a power series 
    \begin{equation}    W(\Phi)=2\Phi+e^{-2\pi\Phi}\sum_{m=0}^\infty \frac{(-1)^m}{(2m)!}\Phi^{2m}\int_0^\infty \rmd \ell \mu(\ell)\;\ell^{2m}\label{eq:dilaton_pot_series},
    \end{equation}
which makes it clear that the weight determines the coefficients of the expansion. In particular, for critical JT gravity, we compute these coefficients using the form of the non-generic critical weight we derived in Subsection \ref{subsection:hyp_o(n)}, $\mu_{\alpha}(\ell)= t_{crit}\;e^{-\ell \sqrt{E_\mu}}\ell^{-\alpha}  \Theta(\ell-\epsilon)$ with $\epsilon>0$ a cutoff we use to ensure that $\mu_\alpha(\ell)<\infty$ for all $\ell$. We get that 
\begin{equation}
    \int_0^\infty \rmd \ell \mu_\alpha(\ell)\;\ell^{2m}=t_{crit}\;\epsilon ^{-\alpha +2 m+1} E_{\alpha -2 m}\left(\sqrt{E_\mu} \epsilon \right),\label{eq:dilaton_coeffs}
\end{equation}
where $E_\gamma(x)$ is the generalized exponential integral function. Since the second term of the dilaton potential \eqref{eq:dilaton_pot_series} rapidly becomes subdominant for $\Phi>1$, due to the exponential suppression, the smallest values of $m$ determine the general features of the potential.
Figure \ref{fig:dilaton_potential_coeff} (left) shows the first coefficient ($m=0$) of this potential for different values of $\epsilon$. In particular, when the cutoff is removed ($\epsilon=0$), this term becomes $\left(\sqrt{E_\mu}\right)^{\alpha-1}\Gamma\left(1-\alpha\right)$, which diverges at $\alpha=2$. This is intriguing since we hypothesize that $\alpha=2$ corresponds to a phase transition of the hyperbolic $O(n)$ model. The plot of the first three coefficients of the dilaton potential in this scenario (Figure \ref{fig:dilaton_potential_coeff}, right) reveals that only the term for $m=0$ exhibits discontinuity at $\alpha=2$. Upon further examination of \eqref{eq:dilaton_coeffs} with $\epsilon=0$, it can be proven that all coefficients $m>0$ remain continuous.
\\
We will further analyze the case $\epsilon=0$, in this case, the dilaton potential is
\begin{align}
    W(\Phi) &=2\Phi
    +t_{crit}\; e^{-2\pi\Phi}\sum_{m=0}^\infty \frac{(-1)^m}{(2m)!}\Phi^{2m} \left(\sqrt{E_\mu}\right)^{\alpha-2m-1}\Gamma\left(2m+1-\alpha\right),
\label{eq:dilaton_pot_critical_particular}
\end{align}
and it is shown in Figure \ref{fig:dilaton_potential} for different values of $\alpha$ and in Figure \ref{fig:dilaton_potential_t} for different values of $t_{crit}$ for the dilute ($\alpha<2$) and dense ($\alpha>2$) phases. \\
It should be emphasized that in the case $\epsilon=0$ the weight $\mu_{\alpha}(\ell)= e^{-\ell \sqrt{E_\mu}}\ell^{-\alpha}$ is not finite at $\ell=0$, so it is not an admissible weight in the probabilistic sense. But, as we will show in the following subsection, this is an interesting thermodynamical case that further relates this non-generic critical family of models with FZZT and EOW branes, in the latter case, the weight \eqref{eq:eow_weight} also diverges at $\ell=0$.
\\
\subsection{Thermodynamics}
\begin{figure}[htbp]
\centering
\includegraphics[trim={0 5cm 0 0},clip,width=\textwidth]{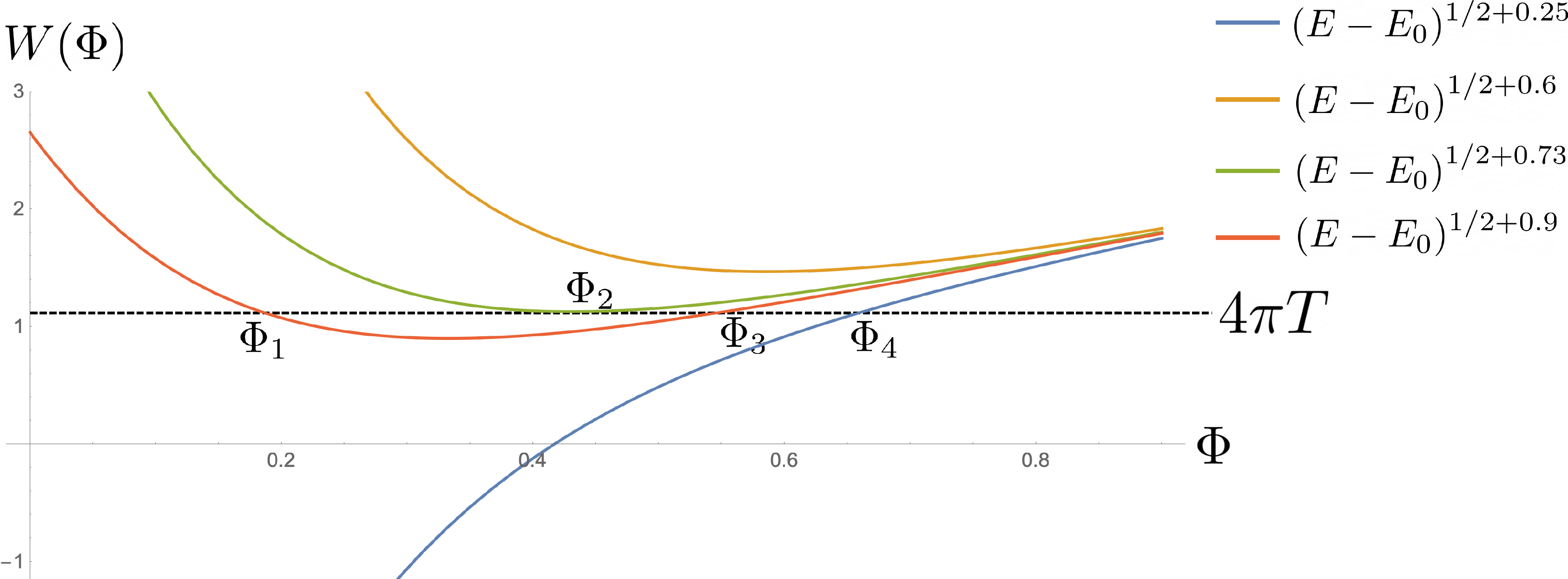}
\caption{The potential \eqref{eq:dilaton_pot_critical_particular} for $\alpha=1.75$ (blue), $\alpha=2.1$ (yellow), $\alpha=2.23$ (green) and $\alpha=2.4$ (red), and fixed values $t_{crit}=5$ and $E_\mu=0.1$. A fixed temperature $T$ is shown as a dotted line. The black hole solutions at temperature $T$ are denoted by $\Phi_1$, $\Phi_2$, $\Phi_3$ and $\Phi_4$ labelled by increasing size.   \label{fig:dilaton_bh}}
\end{figure}

Following the steps of \cite{Witten:2020ert}, we can write the classical euclidean black hole solution to the equations of motion of \eqref{eq:dilaton_action} in the gauge $\Phi(r)=r$,
      \begin{equation}
       ds^2=A(r)^2dt^2+\frac{1}{A(t)^2}dr^2, \hspace{12mm} A(r)=\int_{r_h}^r \rmd x W(x)
      \end{equation}
where $r=r_h$ is the position of the euclidean horizon. To have a well-defined exterior solution, we need this function to be positive $A(r)>0$ for $r>r_h$ and vanish at the horizon. Expanding around the horizon one gets $A(r)=W(\Phi_h)(r-r_h)+\mathcal{O}((r-r_h)^2)$, therefore the latter condition translates into $W(\Phi_h)>0$. Substituting this expansion in the metric and demanding smoothness at $r=r_h$, one obtains the euclidean temperature of the black hole 
\begin{equation}
    T(\Phi_h)=\frac{W(\Phi_h)}{4\pi}.
\end{equation}
Additionally, it can be observed that the energy difference between two black holes with horizon radii $\Phi(r_1) = \Phi_1$ and $\Phi(r_2) = \Phi_2$ is
\begin{equation}
    \Delta E=\frac{1}{2}\int_{\Phi_1}^{\Phi_2}\rmd \Phi \ W(\Phi).
\end{equation}
Thermodynamically speaking, a system is stable if its heat capacity is positive, i.e.\ $\frac{dE}{dT}>0$, which implies that the stability of the euclidean black hole solution with a horizon at $\Phi_h$ in the canonical ensemble ($T$ fixed) requires that $W'(\Phi_h)>0$. In the case of our potential \eqref{eq:dilaton_pot_critical_particular}, it is monotonically increasing in the dilute phase ($3/2<\alpha<2$) and convex in the dense phase ($2<\alpha<5/2$), as shown in Figure \ref{fig:dilaton_potential}. Qualitatively similar behavior has been observed in \cite{Okuyama:2021Page}, but in that case, one regime corresponds to FZZT branes and the other to anti-FZZT branes and the density of states changes abruptly from $\sqrt{E-E_0}$ to $(E-E_0)^{3/2}$. On the other side, in \cite{Gao:2021uro}, where they observe the same change in $\rho_0(E)$ ($\sqrt{E-E_0}$ to $(E-E_0)^{3/2}$) for EOW branes, the dilaton potential is always convex. In our case, this family of non-generic critical models continuously interpolates between these two regimes, and this change is directly linked to the varying scaling of $\rho_0$. \\
To exemplify the thermodynamics of this family of black holes, we showcase four instances of the potential \eqref{eq:dilaton_pot_critical_particular} for different values of $\alpha$ in Figure \ref{fig:dilaton_bh}. We start by fixing a temperature $T$. The first observation is that in the dilute phase, there is only one black hole solution and it is stable. On the other side, in the dense phase, there are either no black hole solutions (yellow graph), one thermodynamically unstable solution (green), or two black holes (red), one unstable ($\Phi_1$) and one stable ($\Phi_3$). Therefore, in this case, the only two stable black holes at temperature $T$ are $\Phi_3$ and $\Phi_4$. As it was noted in \cite{Witten:2020ert}, the stable solution that minimizes the free energy of the system is the one with the largest horizon, which in this case corresponds to $\Phi_4$. In other words, given a fixed temperature and a fixed value for $t_{crit}$, there is a thermodynamically preferred value of $\alpha$ and, therefore, in the brane picture, a preferred geometry, either dense or dilute (see Figure \ref{fig:holes_JT}), the latter is the case for this example.\\   
\\
Another remark we want to make is that the two thermodynamical conditions of the potential, $W(\Phi_h)>0$ and $W'(\Phi_h)>0$, define the physical consistency requirements for any non-generic critical potential we can construct, which are additional constraints for the form of the weights $\mu_\alpha(\ell)$. So, if we start with a general potential \eqref{eq:dilaton_pot_general} without assuming the exact form of $\mu_\alpha$ and want to construct thermodynamically stable black hole solutions, it implies that
\begin{equation}
     0<\exp^{-2\pi \Phi_h}\int_0^\infty \rmd \ell \ \ell\mu_{\alpha}(\ell)\sin{(\ell\Phi_h)}< 2+4\pi\Phi_h
\end{equation}
needs to be satisfied. It is to be remarked that this inequality admits weights that diverge at $\ell=0$, even though these are not allowed if we want a probabilistic interpretation of non-generic critical JT gravity. Furthermore, in terms of the physical interpretation of this family of non-generic critical models, they can be interpreted as a family of thermodynamic stable brane solutions with finite entropy, to which evaporating black holes can decay. However, it is important to note that the rigorous derivation of these branes from String Theory is still an open question.

\section{Spectral form factor}\label{section:SFF}
\begin{figure}[htbp]
\centering
\includegraphics[width=.5\textwidth]{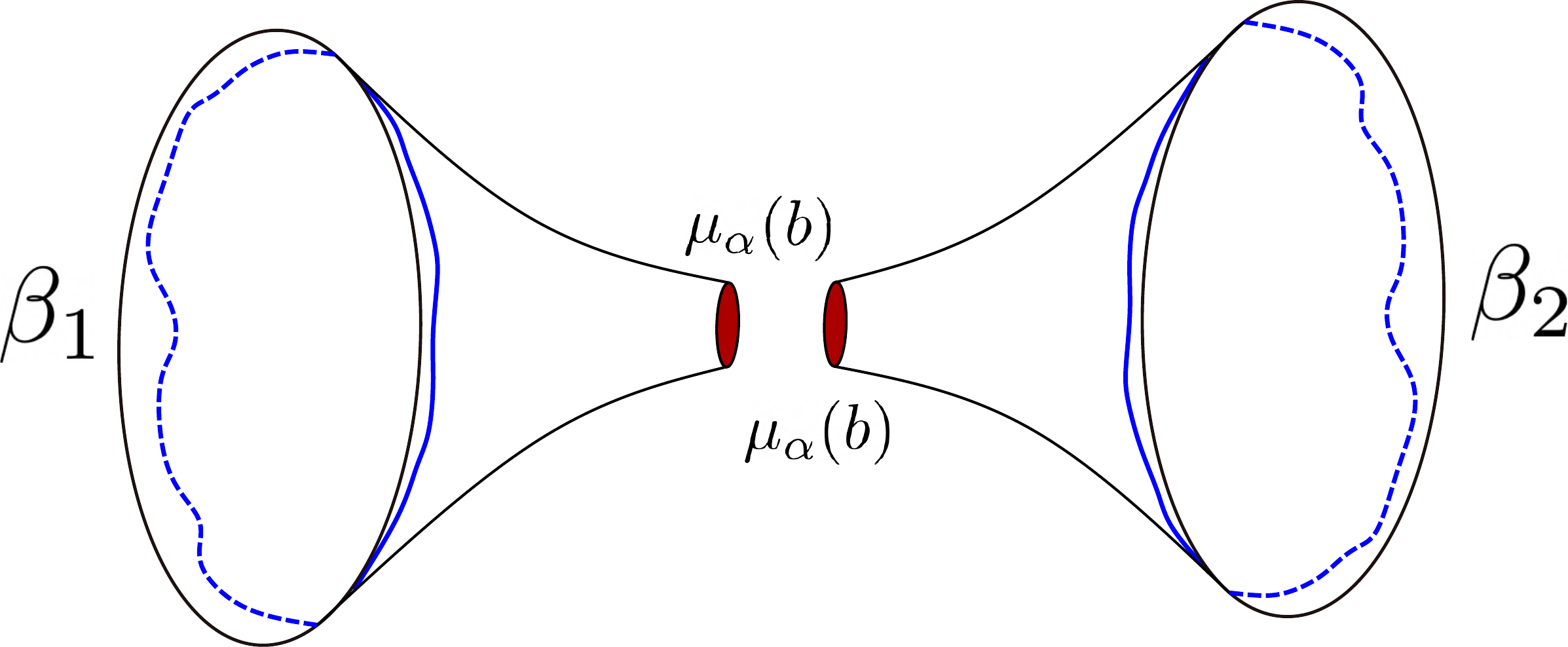}
\caption{Double trumpet contribution arising from gluing two half-wormholes ending each on a brane with non-generic critical weight.    \label{fig:double_trumpet}}
\end{figure}

\begin{figure}[htbp]
\centering
\includegraphics[width=.9\textwidth]{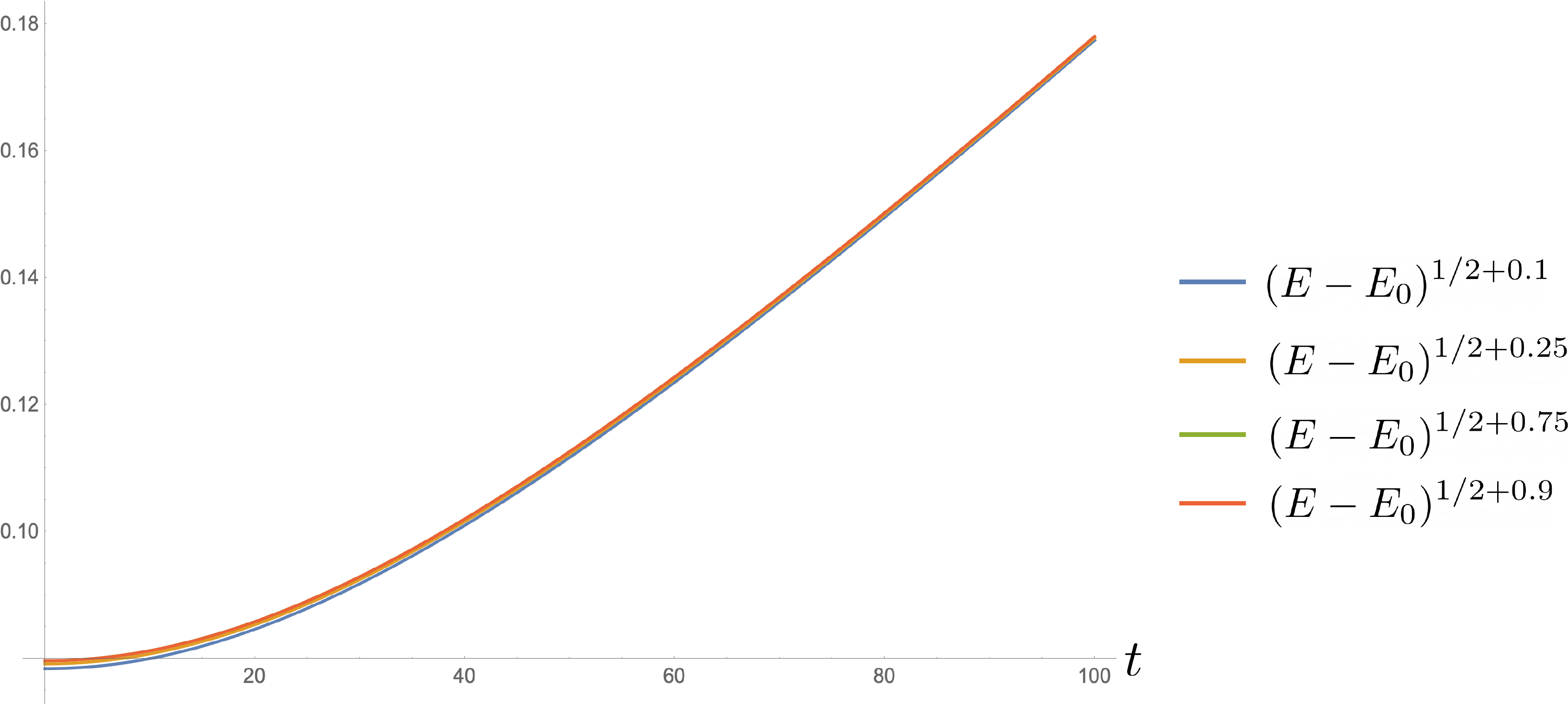}
\caption{Spectral form factor \eqref{eq:sff} for different values of $\alpha$, $t_{crit}=1$, $\beta=50$ and $E_\mu=0.1$.    \label{fig:sff}}
\end{figure}

The spectral form factor in JT gravity is given by the time-ordered two-point correlation function, which in geometrical terms corresponds to the double trumpet for genus zero and no defects. 
In matrix models, it displays distinctive and universal characteristics, notably the presence of a ramp followed by a plateau structure. Interestingly, in JT gravity the spectral form factor exhibits a ramp-like behavior without a well-defined plateau.
The absence of a plateau is a consequence of the geometric structure of the double trumpet, whose dominant contribution leads to a continuous growth of the spectral form factor without reaching a plateau. The absence of a plateau in the spectral form factor indicates that the dual CFT does not possess a well-defined time scale for thermalization.\\
In earlier studies (e.g. \cite{Okuyama:2021FZZT}), it has been investigated if a plateau can be recovered with the inclusion of branes. Hence, it arises as a natural question to investigate it using our configuration of branes.
In our case, the main contribution to the double trumpet corresponds to two trumpets (or half-wormholes) glued at geodesic boundaries, each with non-generic critical weight $\mu_{\alpha}(\ell)= e^{-\ell \sqrt{E_\mu}}\ell^{-\alpha}$ (see Figure \ref{fig:double_trumpet}). This corresponds to
\begin{align}
    Z_{0,2}(\beta_1,\beta_2) &= \int_0^\infty \rmd b\left(\mu_\alpha(b)\frac{e^{-b^2/4\beta_1}}{2\sqrt{\pi \beta_1}}\right) \left(\mu_\alpha(b)\frac{e^{-b^2/4\beta_2}}{2\sqrt{\pi \beta_2}}\right)\\
    &=   \frac{t_{crit}}{4\pi}\; \frac{\Gamma (1-\alpha ) \left(\frac{\beta_1\beta_2}{\beta_1+\beta_2}\right)^{\frac{1-\alpha }{2}} U\left(\frac{1}{2}-\frac{\alpha }{2},\frac{1}{2},E_\mu\frac{ \beta_1 \beta_2}{\beta_1+\beta_2}\right)}{ \sqrt{\beta_1\beta_2}},
\end{align}
where $U$ is the confluent hypergeometric function. Therefore, the spectral form factor is 
\begin{equation}
    Z_{0,2}(\beta+it,\beta-it)=\frac{t_{crit}}{\pi}\;\frac{2^{\alpha -\frac{5}{2}} \Gamma (1-2 \alpha ) \left(\beta +\frac{t^2}{\beta }\right)^{\frac{1}{2}-\alpha } U\left(\frac{1}{2}-\alpha ,\frac{1}{2},\frac{2E_\mu \left(t^2+\beta ^2\right)}{\beta }\right)}{ \sqrt{\beta ^2+t^2}}.\label{eq:sff}
\end{equation}
This is shown in Figure \ref{fig:sff}, where we can observe that the spectral form factor grows linearly without reaching a plateau. Regardless, this is a computation at $g=0$ and $\hbar\rightarrow 0$, so the possibility that these and more non-perturbative corrections change the late time behavior or \eqref{eq:sff} cannot be ruled out. It is also worth noting that the low temperature and late time ($\beta\gg 1$ and $t\gg 0$) behavior of the spectral form factor does not depend on the values of $\alpha$.     
\section{Discussion and outlook}\label{Section_Discussion}
Probably the most important result of this paper is to show how JT gravity and random geometry are deeply connected and how tools from both fields are useful for computational goals but also for our conceptual understanding. In this paper, we explored the random geometry aspects of JT gravity, offering insights into its critical regimes. 
Furthermore, we have remarked once again `the power of the string equation', already highlighted in multiple works \cite{Johnson:2019eik,Johnson:2021tnl,Turiaci:2020fjj}, noting its capacity not only to compute partition functions and density of states at the leading order but as a translation tool between random geometry and JT gravity. \\
By using tools coming from the study of Boltzmann planar maps, we uncovered what we call critical phases of JT gravity, which bear resemblance to phase transitions observed in planar maps, from surfaces with microscopic holes in the subcritical regime, to those with macroscopic ones in the non-generic critical regime. The occurrence of such transitions is the result of fine-tuning the weights of geodesic boundaries, which can be thought of as brane configurations. Furthermore, we made this analogy precise by determining the relation between hyperbolic surfaces with non-generic critical weights with $3/2<\alpha<5/2$ and the hyperbolic $O(n)$ loop model with $n=2\sin\left(\pi\left(\alpha-\frac{1}{2}\right)\right)$. We showed that the different values of $\alpha$ significantly impact the density of states of the system. Through our analysis, we have identified a family of models that smoothly interpolate between the density of states behaviors of $\rho_0(E)\sim\sqrt{E-E_0}$ and $\rho_0(E)\sim (E-E_0)^{3/2}$, thereby establishing connections to JT gravity coupled to dynamical EOW branes \cite{Gao:2021uro} and FZZT branes \cite{Okuyama:2021Page}. Moreover, we showed in which strict sense this transition from $\rho_0(E)\sim\sqrt{E-E_0}$ to $\rho_0(E)\sim (E-E_0)^{3/2}$ in the density of states is a proper phase transition in the physical sense and in which sense it is different compared to other transitions that have been identified previously \cite{Johnson:2020lns}. This is when the expected number of geodesic boundaries, or defects, diverge. Additionally, by analysing the effective dilaton potential for these models, one can distinguish between two subsets of models, $\alpha>2$ and $\alpha<2$ which suggests another phase transition at $\alpha=2$.\\
In light of our results, there are several interesting paths for future research to explore. 
Firstly, it remains essential to address the non-perturbative stability of the identified critical regimes and their compatibility with alternative definitions of EOW branes \cite{Johnson:2021tnl}.
This is related to an important question that arises, whether there exists a matrix model whose double-scaled density of states follows the form $(E-E_0)^{\alpha-1}$. Such a matrix model was constructed in \cite{Ambjorn:2016lkl}, but its non-perturbative study poses a big challenge due to the infinite order of its KdV differential equations \cite{Ambjorn:2017ibv}. We plan on addressing this and investigating the properties of such a matrix model.
\\
Additionally, from the mathematical point of view, exploring the phase transition at $\alpha=2$ in the hyperbolic $O(n)$ loop model is an open problem. Our conjecture is that a transition from non-touching geodesic boundaries to overlapping ones should occur similarly to the planar map case. This line of work may help to better understand the geometries in Figure \ref{fig:holes_JT} and their physical meaning either from a fluctuating or from a decorated hyperbolic geometry point of view in a similar way as for planar maps.\\
Furthermore, we have presented a way to couple JT gravity to any kind of (admissible) defects using the string equation. This allows us to compare different results in the JT gravity literature in a single and precise context. For example, in \cite{Blommaert:2021fob}, the weights are chosen to achieve factorization in JT gravity. Such weight, $\mu(\ell_1,\ell_2)=-\frac{1}{\ell_1}\delta(\ell_1-\ell_2)$, is bilocal and negative valued which is not compatible with a probabilistic interpretation, but allows to `unglue' the double trumpet allowing for the factorization of the two boundary amplitude $\langle Z(\beta_1) Z(\beta_2)\rangle$. An even more interesting case for us is the weight choice used in \cite{Blommaert:2021gha} to obtain a discrete JT gravity spectrum. In this context, our non-generic critical weights are equivalent to finding a background matrix $H_0$ such that $\mathrm{Tr}(H_0)^k\sim k^{-\alpha}$ for $k\gg 1$. In this regime, when big holes appear in the hyperbolic surface, the density of states $\rho_0$ is conjectured to become discrete. Nevertheless, our results for $\rho_0$ do not agree with it, its leading order behavior changes from a square root in a non-trivial way but the spectrum stays continuum for all $\alpha$. The incompatibility of these two results poses an interesting avenue for further research. The last case we want to comment on is \cite{Okuyama:2021Page}, where the weights of anti-FZZT branes were tuned to reproduce a Page-like curve which exactly corresponds to a transition from $\rho_0(E)\sim\sqrt{E-E_0}$ to $\rho_0(E)\sim (E-E_0)^{3/2}$. Since our models interpolate between these two regimes, we think that the exploration of the non-generic critical phase raises intriguing questions regarding the implications for the Page curve and black hole evaporation.\\
Finally, we want to make a comment on the significance of non-generic critical models from the holographic point of view. One could forget about the geometry of the bulk and only look at the density of states of the quantum system at the holographic boundary. Then, in the subcritical regime $\rho_0(E)\sim\sqrt{E-E_0}$, which is the leading order behavior of the Schwarzian. However, in the non-generic critical regime, the density of states behaves as $\rho_0(E)\sim (E-E_0)^{\alpha-1}$, so a question that arises is if the transition from a subcritical to a non-generic critical regime can be put in the framework of deformations of the Schwarzian or if there are quantum systems that naturally present this the non-generic critical behavior, but we leave this for future work.

\acknowledgments
I am grateful to Timothy Budd, Clifford Johnson, Bart Zonneveld and Pjotr Koster for valuable discussions and comments on a draft. In particular, Timothy Budd contributed to the analysis of the hyperbolic $O(n)$ loop model in Section \ref{section:Random_geometry}. This work is part of the START-UP 2018 programme with project number 740.018.017, which is financed by the Dutch Research Council (NWO).


\bibliographystyle{JHEP}
\bibliography{biblio.bib}

\end{document}